\documentclass{cernrep} 
\usepackage{texnames}
\usepackage[T1]{fontenc}
\begin{document}
\title{An introduction to cosmology}
 
\author{Kerstin E. Kunze}

\institute{Departamento de F\'\i sica Fundamental and IUFFyM,
Universidad de Salamanca, Plaza de la Merced s/n, 37008 Salamanca,
Spain}

\maketitle % this produces the title block

\begin{abstract}
Cosmology is becoming an important tool to test particle physics models.
We provide an overview of the standard model of cosmology with an emphasis 
on the observations relevant for testing fundamental physics.
\end{abstract}
 
\section{Introduction}
\label{sec1}
\setcounter{equation}{0}

Cosmology is the only part of physics which has the whole universe as its area of research.
As such it covers a vast range of scales.
Energy scales go from the present day temperature of  $10^{-4}$ eV upto the 
Planck scale $10 ^{19}$ GeV. It aims to describe the evolution of the universe from its very beginning  upto today 
where it has an estimated age of the order of $10^{10}$ years.
Due to its very nature of understanding the universe as a whole  cosmology needs input from
very different areas of physics. These naturally include astrophysics and theories of gravitation,
but also plasma physics, particle physics and experimental physics.

Over the last two decades cosmology has entered a data driven era. A turning point were the first 
observations with upto then unprecedented precision of the cosmic microwave background (CMB)
with the COBE satellite in 1990 \cite{firas1,firas2,firas3}. 
Since then there have been a variety of  CMB experiments, ground based such as the Atacama Cosmology telescope (ACT) \cite{act} in Chile,
the South Pole telescope (SPT) \cite{spt} at the South Pole, with balloons such as Boomerang \cite{boomerang}, two more
satellites, namely the Wilkinson Microwave Anisotropy Probe (WMAP) \cite{wmap} and Planck \cite{planck} and more experiments are planned for the 
future. 
The first systematic study of the  structure in the local universe was the Center for Astrophysics (CfA) survey of galaxies \cite{cfa}. One of the most recent ones is the Sloan Digital Sky Survey III \cite{sdss-3} with SDSS IV \cite{sdss-4} already underway.
In 2019 the launch of the  ESA mission EUCLID is planned. Its goal is to  measure shapes and  redshifts of galaxies upto redshifts of $z\sim 2$ thereby allowing to
determine the evolution of the recent universe since the time when dark energy became important \cite{euclid}.

In these lectures we will start with the evolution of the universe on very large scales where it is, to a high degree, isotropic and homogeneous. This will be followed by 
a description of the thermal history of the universe from very early times upto the present epoch including the key events.
Observations show that the universe is not perfectly isotropic. At the largest scales this manifests itself in the small temperature anisotropies $\frac{\Delta T}{T}\sim {\cal O}(10^{-5})$
 in the CMB. These are an imprint of the density perturbations which provided the seeds from which all large scale structure such as galaxies have developed.
 Therefore the second part of the lectures is dedicated to the inhomogeneous universe, the origin of the temperature anisotropies and polarization of the 
 CMB and large scale structure.
 The third and last part deals with the two big unknowns in our universe. Different observations such as from the CMB, large scale structure and high redshift supernovae
 pinpoint the cosmological parameters to around 4\% baryonic matter,  25\% of cold dark matter and about 71\% of dark energy. 
From the data the physical properties of these components can be constrained. However, upto the present it is neither known what constitutes dark matter nor dark energy. There are many proposals, some of which are rather exotic, but none stands out as a ''natural'' model.

There are already  a number of excellent text books which cover different aspects of these lectures, e.g., \cite{Lyth:2009zz}\cite{durrer},\cite{LMMP},\cite{Coles:1995bd}, \cite{Perkins:2003pp},\cite{PU}, \cite{KT}, \cite{mvdbw}. 
 Also the review sections related to cosmology in \cite{PDG} provide a very good overview.
 
 \section{The homogeneous universe}
 \label{sec2}

 Observations such as the high degree of isotropy of the CMB indicate that globally the universe is well described by a spatially homogeneous and isotropic model. These are  the Friedmann-Robertson-Walker solutions of general relativity to which a brief introduction can be found in  appendix \ref{A1}.  Spatial homogeneity and isotropy mean that physical conditions are the same everywhere and in each direction. It allows to choose a coordinate system such that the four dimensional space time is described by a foliation of spatial hypersurfaces at constant time and the metric is given by
 \begin{eqnarray}
ds^2=-dt^2+a^2(t)\left[\frac{dr^2}{1-kr^2}+r^2\left(d\theta^2+\sin^2\theta d\phi^2\right)\right].
\end{eqnarray}
The parameter $k$ labels the different choices of spatial curvature.
It takes the values $k=0$ for a flat universe, $k=+1$ for a closed universe and $k=-1$ for an open universe.
The coordinates $r$, $\phi$ and $\theta$ determine the spatial {\it comoving} coordinates on each constant time slice. These coordinates do not change during the evolution of the universe or, in other words, from one constant time slice to the next one. 
However, as the universe is not static  this leads naturally to the notion of {\it physical} coordinates as well as {\it physical}  scales. As an example,
consider two nearby observers (or galaxies or any other astrophysical object) and assume that at some fixed time $t_1$ they are separated by a  distance $\ell_1$. 
Because of the expansion of the universe all {\it physical} scales are multiplied by the scale factor $a(t)$ so that $\ell_1=a(t_1)\ell_{com}$.
Hence, at some later time $t_2$ the {\it physical} distance between the two objects is given by 
$\ell_2=a(t_2)\ell_1/a(t_1)$. Quite often the scale factor today is set to one, $a(t_0)=1$, in which case 
at present comoving and physical scales coincide \footnote{Note, however, that in this case the curvature parameter $k$ has to be appropriately rescaled in the case of a non flat universe.}. Here we have introduced another common choice, i.e. to denote the present epoch by an index "0". Choosing the coordinate system accordingly the physical radial distance is given by
\begin{eqnarray}
d_p=\int_0^r\frac{a dr'}{(1-kr^{'\,2})^{\frac{1}{2}}}\equiv a(t)f(r).
\end{eqnarray} 
Thus the physical distance between two objects changes locally at a rate $v_p=\dot{a}f(r)=\frac{\dot{a}}{a}d_p(t)=H(t)d_p(t)$, where $H\equiv\frac{\dot{a}}{a}$ is the Hubble parameter. In this section a dot indicates the derivative w.r.t. cosmic time $t$. 
Applied to the present epoch $v_p=H_0 d_p$ which is also known as Hubble's law.
In an expanding universe this is a recession velocity which was first observed in galaxies  by Edwin Hubble in 1929.  Present observations give a value of the Hubble constant close to 70 km s$^{-1}$Mpc$^{-1}$. There are some variations in the value of $H_0$ depending on which data are used resulting in a certain tension between different data sets. For example, from observations of supernovae in combination with Cepheid variables
$H_0=(73.8\pm 2.4)$ km s$^{-1}$Mpc$^{-1}$ \cite{Riess:2011yx} and from the Planck 15 temperature data combined with the Planck 15 gravitational lensing reconstruction the Hubble parameter is found to be $H_0=(67.8\pm0.9)$ km s$^{-1}$Mpc$^{-1}$ \cite{planck15-cosmo}.

\begin{figure}
\centering{\includegraphics[width=0.7\linewidth]{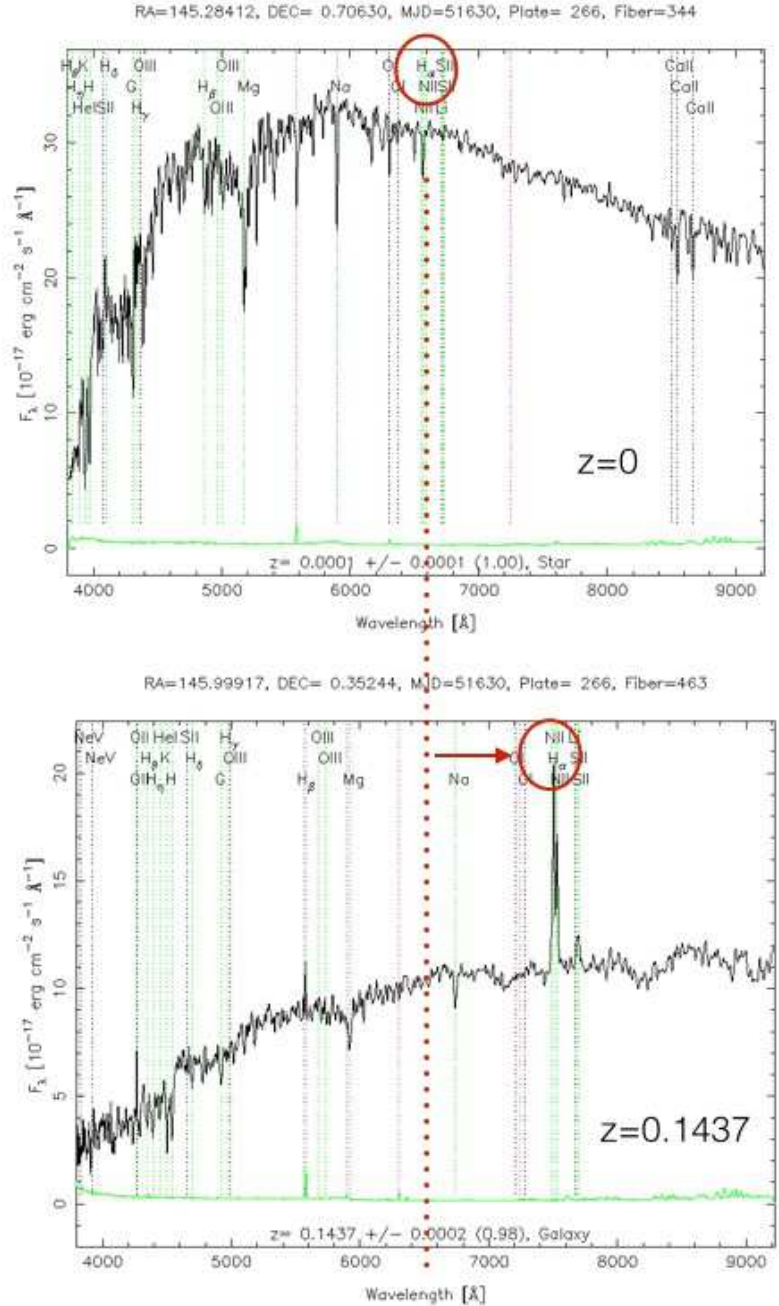}}
\caption{Observing the cosmological redshift. {\it Upper panel:} A reference spectrum of a star at redshift $z=0$. {\it Lower panel:} Spectrum of galaxy SDSS J094359.79+002108.7  at redshift $z=0.1437$. Marked by a red circle is in both spectra the H$_{\alpha}$ line of the Balmer series of hydrogen ($\lambda=6564.7$\AA  $\;$at rest, i.e. at $z=0$). The red arrow in the lower panel indicates the shift of the H$_{\alpha}$ line (as well as the other spectral lines) towards the long wavelength part of the spectrum due to the expansion of the universe, implying that in this case a cosmological redshift is observed.  
Whereas in the case of the reference spectrum ({\it upper panel}) the line is an absorption line in the case of the spectrum of the galaxy ({\it lower panel}) it is an emission line. The figures have been done using spectra taken from Data Release 7 (DR7) of the Sloan Digital Sky Survey, SDSS-II \cite{sdss-dr7}. The spectra have been obtained from http://skyserver.sdss.org/dr7/en/tools/.}
\label{fig:fig1.eps}
\end{figure}

An important question is how this recession velocity can actually be measured. The answer lies with the observation of cosmological redshift.  Consider a distant galaxy at radial coordinate $r_1$ which emits light at some time $t_1$ at, say, wavelength $\lambda_e$ which is observed by an observer at $r=0$ at the present time $t_0$ at a wavelength $\lambda_o$. Light travels along null geodesics so that $ds^2=0$ implying 
 for radial null geodesics
 \begin{eqnarray}
 \int_{t_1}^{t_0}\frac{dt}{a(t)}=\int_0^{r_1}\frac{dr}{(1-kr^2)^{\frac{1}{2}}}=f(r_1).
 \end{eqnarray}
 Light emitted at a time $t_1+\delta t_1$ reaches the detector at a time $t_0+\delta t_0$. Since $f(r_1)$ is a constant and it is assumed that the source has no peculiar motion,
 \begin{eqnarray}
 \int_{t_1}^{t_0}\frac{dt}{a(t)}=\int_{t_1+\delta t_1}^{t_0+\delta t_0}\frac{dt}{a(t)}\Rightarrow 
 \int_{t_1}^{t_1+\delta t_1}\frac{dt}{a(t)}=\int_{t_0}^{t_0+\delta t_0}\frac{dt}{a(t)}.
 \end{eqnarray} 
 Thus for small $\delta t_i$, $i=0,1$, ($\lambda c\delta t_i\ll ct_i$) and approximating $a(t)\sim\; const.$ during the time interval of integration it follows that $\delta t_1/a(t_1)=\delta t_0/a(t_0)$ so that 
 \begin{eqnarray}
1+z\equiv \frac{\lambda_0}{\lambda_1}=\frac{a(t_0)}{a(t_1)}
 \end{eqnarray}
 where $z$ is the redshift. This cosmological redshift is a direct consequence of the expansion of the universe. 
 In case, the scale factor is diminishing a blue shift is observed.  In \Fref{fig:fig1.eps} an example of the observation of cosmological redshift in the spectrum of a galaxy at redshift $z=0.1437$ together with a reference spectrum of a star at $z=0$ is shown.
 
\begin{figure}
\centering{\includegraphics[width=0.3\linewidth]{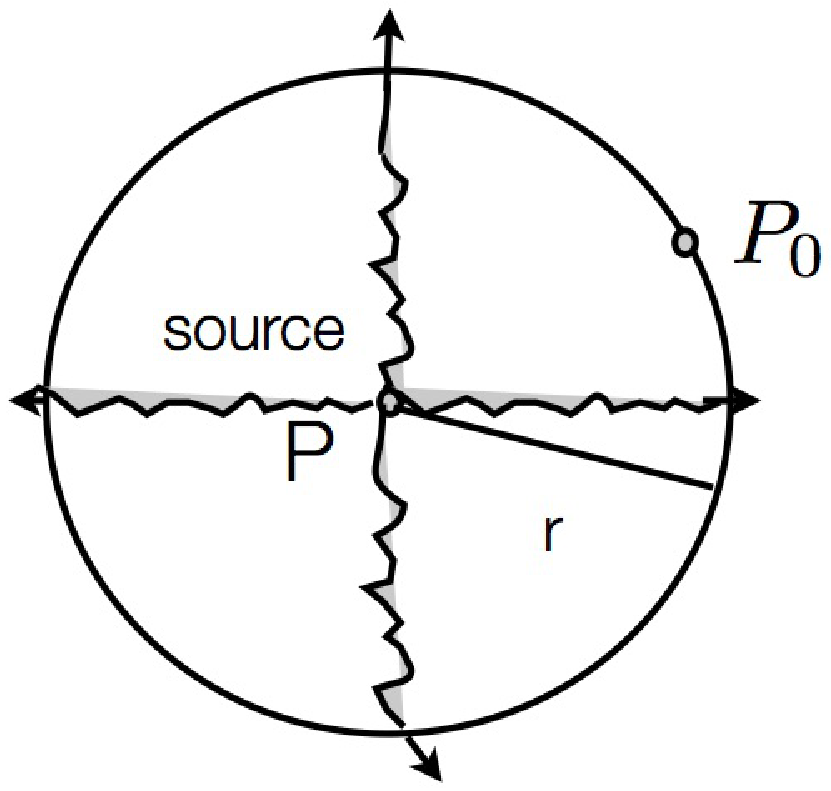}}
\caption{A point source situated at $P$ is emitting radiation isotropically which is received by an observer at $P_0$ at a coordinate distance $r$. Knowing the absolute luminosity of the point source allows the observer to determine its luminosity distance $D_L$. }
\label{fig:fig2.eps}
\end{figure}

It is difficult to measure distances at very large scales even more  so because of the expansion of the universe. 
In cosmology there are  two important distance measures which are the luminosity distance and the angular diameter distance. These rely on the fact that the flux and the angular size of an object could in principle be known.
This means that if the observer has independent knowledge of 
 its absolute luminosity or its physical size  its distance can be estimated. Objects whose absolute luminosity or physical size are available define the classes of standard candles or standard rulers, respectively. Distances obtained for these objects are the luminosity distance and angular diameter distance, respectively.
The luminosity distance $D_L$ is defined by 
\begin{eqnarray}
D_L=\left(\frac{L}{4\pi\ell}\right)^{\frac{1}{2}},
\end{eqnarray}
where $L$ is the absolute luminosity and  $\ell$ is the visible luminosity which is  received by the observer.
To determine its evolution with redshift consider a source located at a point $P$ at a coordinate distance $r$ and emitting a signal at some time $t$,
as illustrated in \Fref{fig:fig2.eps}.
An observer at a point $P_0$ observes the signal at a time $t_0$. The absolute luminosity is the energy flux (=energy/time) and the visible luminosity $\ell$ is the energy flux density (=energy/(time$\times$surface)).
The rates of emission and reception of photons are related by $(\delta t_0)^{-1}/(\delta t_e)^{-1}=a(t)/a_0$. Thus the absolute luminosity $L=E_e/\delta t_e=hc/\lambda_e\delta t_e$ and the visible luminosity $\ell=E_0/(4\pi a_0^2r^2\delta t_0)=hc/(4\pi a_0^2r^2\lambda_0\delta t_0)$ resulting in 
\begin{eqnarray}
D_L=a_0r(1+z).
\label{eq:D_L}
\end{eqnarray}
Hubble's law implies locally that  an object moves away from the observer with a velocity proportional to its physical distance. Expanding the scale factor beyond linear order leads to
\begin{eqnarray}
D_L\simeq H_0^{-1}\left[z+\frac{1}{2}(1-q_0)z^2+...\right]
\end{eqnarray}
 where $q_0$ is the deceleration parameter today which in general is defined by $q\equiv -\frac{a\ddot{a}}{\dot{a}^2}$.
In order to determine the luminosity distance it is necessary to find objects with a known absolute luminosity  which can be used as standard candles. Cepheids provide one example of standard candles which have been used frequently in the past.
These are variable stars with a regular change in their apparent magnitude whose period is related to their absolute luminosity. 
Another class of standard candles are Type Ia supernovae (SN Ia) which have a well known light profile (cf. e.g \cite{Perlmutter}, \cite{Schmidt:1998ys}).
They are found in binary systems formed when one of the stars is accreting material from the other star. 
Reaching a critical mass limit leads to a thermonuclear explosion and subsequently to a sudden increase in the observed
light curve. These light curves are well known and can be observed at cosmological distances as they can reach luminosities upto $10^{10}$ times the luminosity of the sun. The observations of SN Ia have been particularly important in establishing that the expansion of the universe is accelerating in the current epoch. 

\begin{figure}
\centering{\includegraphics[width=0.7\linewidth]{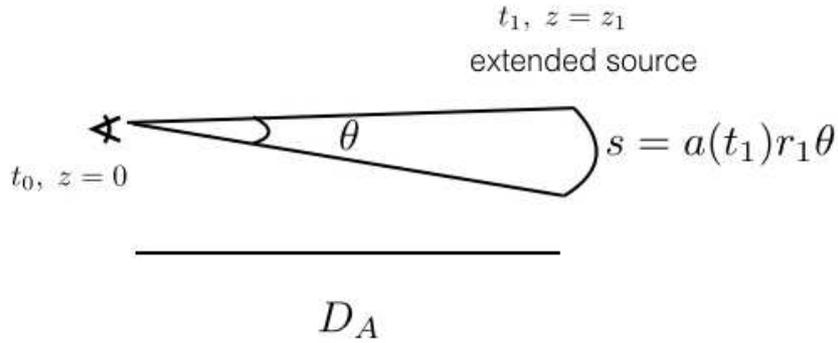}}
\caption{The angular diameter distance $D_A$ is defined in such a way that $\theta=s/D_A$ where $s=a(t_1)r_1\theta$ is the physical extension of the object. }
\label{fig:fig3.eps}
\end{figure}

The other important distance measure is the angular diameter distance $D_A$ which generalizes the concept of the parallax. It is defined by requiring that the angle $\theta$ over which the object extends is given, just as in Euclidean space, that is, by the ratio of the transverse size $s$ over the distance $D_A$ to the object (cf \Fref{fig:fig3.eps}). 
Assuming that the extended source at a redshift $z_1$ is emitting light at a time $t_1$ and is located at a comoving radial distance $r_1$ then for small angles $\theta$, the angular diameter distance is given by $D_A=a(t_1)r_1$.
Comparing this with the expression of the luminosity distance $D_L$ (cf. Eq. (\ref{eq:D_L})) it is found that 
\begin{eqnarray}
\frac{D_A}{D_L}=(1+z)^{-2}.
\end{eqnarray}
The angular diameter distance $D_A$ is only useful if the angular extension of the source is known which in general
is a difficult task to measure at cosmological distances.
As such it relies on a standard ruler.  As will be seen in the next section such a standard ruler does exist in the early universe and is imprinted in the observed angular power spectrum of the temperature fluctuations of the cosmic microwave background.

\begin{figure}
\centering{\includegraphics[width=0.45\linewidth]{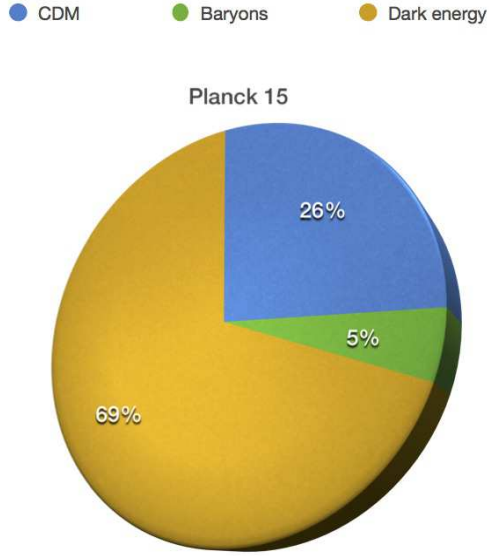}}
\caption{Distribution of contributions to the total energy density at present from Planck 15 best fit parameter. }
\label{fig:fig4.eps}
\end{figure}

 Due to the high degree of symmetry of the Friedmann-Robertson-Walker (FRW) solutions the description of the evolution of the universe becomes particularly simple. The evolution of the background 
 geometry is encapsulated in one time dependent function which is the scale factor $a(t)$. In the physically relevant cases matter is described by a perfect fluid determined by its energy density $\rho(t)$ and pressure $p(t)$.  These are related  by an equation of state which has a rather simple form, $p=w\rho$, where $w$ is a constant. 
 In this case the evolution of the background is described by the Friedmann equations
 \begin{eqnarray}
 H^2+\frac{k}{a^2}&=&\frac{8\pi G}{3}\rho
 \label{eq:friedmann}\\
 \frac{\ddot{a}}{a}&=&-\frac{4\pi G}{3}\left(\rho+3 p\right)
 \label{eq:ddota}
 \end{eqnarray}
 where $H=\frac{\dot{a}}{a}$ is the Hubble parameter.
 The evolution of matter is determined by
 \begin{eqnarray}
 \dot{\rho}+3H(\rho+p)=0.
 \end{eqnarray}
 In the standard model of cosmology the universe is initially very hot and dense before cooling down as it expands.
Thus initially it 
 is dominated by radiation, subsequently by  non relativistic matter and at  present day by an effective cosmological constant, otherwise also called dark energy which will be discussed in more detail in section \ref{sec3}. These different epochs are described by a perfect fluid with equation of state with $w=\frac{1}{3}$ for radiation, $w=0$ for non relativistic matter and $w=-1$ for an effective cosmological constant.
 The energy density scales as $\rho\sim a^{-4}$ for a radiation dominated universe and as $\rho\sim a^{-3}$ for a matter dominated universe.
  In a flat universe ($k=0$) it is found that the scale factor evolves as $a\sim t^{\frac{1}{2}}$ in a radiation dominated universe and as  $a\sim t^{\frac{2}{3}}$ in a matter dominated universe.
An important quantity in cosmology is the  dimensionless density parameter $\Omega$ which is defined by 
 \begin{eqnarray}
 \Omega\equiv\frac{\rho}{\rho_{crit}}
 \end{eqnarray}
 where $\rho_{crit}\equiv\frac{3H^2}{8\pi G}$ is the critical energy density which corresponds to the total energy density of a flat FRW universe. Using this definition the Friedmann equation takes the form 
 \begin{eqnarray}
 \frac{k}{H^2a^2}=\Omega-1.
\label{eq:oK}
 \end{eqnarray}
 Moreover, including the different contributions to the total energy density explicitly, then
 \begin{eqnarray}
 H^2=\frac{8\pi G}{3}\rho_R+\frac{8\pi G}{3}\rho_M-\frac{k}{a^2}+\frac{\Lambda}{3}
 \end{eqnarray}
 which, with the appropriate definitions, can be written as a constraint equation on the sum of the density parameters,
 \begin{eqnarray}
 \Omega_r+\Omega_m+\Omega_{\Lambda}-\Omega_K=1.
 \end{eqnarray}
 Here $\Omega_R=\Omega_{\gamma}+\Omega_{\nu}$ includes radiation (photons) $\Omega_{\gamma}=\rho_{\gamma}/\rho_{crit}$ and relativistic matter such as light neutrinos $\Omega_{\nu}=\rho_{\nu}/\rho_{crit}$. Non relativistic matter $\Omega_m=\Omega_b+\Omega_c$ has contributions from baryons $\Omega_b=\rho_b/\rho_{crit}$ and cold dark matter $\Omega_c=\rho_c/\rho_{crit}$. The cosmological constant term or dark energy is $\Omega_{\Lambda}=\Lambda/3H^2$ and the curvature term $\Omega_K=k/(a^2H^2)$. Best fit values of 
 68\% confidence limits on the present day values of
 the density parameters  from observations of the cosmic microwave background give for the standard 6-parameter $\Lambda$CDM model from the Planck 15 data \cite{planck15-cosmo}: 
 %Table 4
 $H_0=(67.31\pm0.96)$ km s$^{-1}$Mpc$^{-1}$, $\Omega_m=0.315\pm 0.013$, $\Omega_{\Lambda}=0.685\pm0.013$. 
 %Table 5
 Including spatial curvature as a free parameter gives the 95\% limit 
 $\Omega_K=-0.052^{+0.049}_{-0.055}$.
 In \Fref{fig:fig4.eps} the distribution of the main contributions are shown for Planck 15.
In  section \ref{sec3} we will discuss  the nature of these contributions of which only the baryonic contribution is understood which only makes up about $5\%$ of the total energy density in the universe (cf \Fref{fig:fig4.eps}).

The standard model of cosmology describes the evolution from its very early stages upto the present. 
The underlying model is the rather simple flat Friedman-Robertson-Walker model which in a sense is quite remarkable.  
The Friedman-Robertson-Walker solutions for a standard type of matter such as relativistic or non relativistic particles or radiation generally have an initial curvature singularity which is a consequence of  the Penrose-Hawking singularity theorems. In the standard model of  cosmology, also known as the big bang model, the universe evolves from a very 
tiny, very hot initial state to the present day very large and rather cold state as indicated by the observed temperature of the cosmic microwave background of 2.73 K. In the following we will give a brief overview of the milestones in the evolution of the universe (cf., e.g., \cite{Liddle:2000cg}).
Before times marked by the Planck time  $t\sim 10^{-43}$s nothing is really known since general relativity as a classical theory is not valid anymore. At later times, for temperatures below $T<10^{19}$ GeV all particles of the standard model of particle physics constitute the primordial, relativistic plasma. At temperatures 100-300 MeV, at a time of the order of $10^{-5}$ s the quark-gluon phase transition takes place and free quarks and gluons combine to form baryons and mesons.
The next important epoch is primordial nucleosynthesis when the universe has cooled down to temperatures between 
10 and 0.1 MeV, corresponding to times between $10^{-2}$ to $10^{2}$ s. During this time light elements are produced such as hydrogen, helium-4, deuterium, lithium and other light elements. Observations of the primordial  abundances of these elements provide the possibility to test the standard model of cosmology very deep into its very early evolution.
As will be discussed below in  more detail, during this epoch light, standard model neutrinos decouple at around 1 MeV from the rest of the cosmic plasma. At around 1 eV, at a time $10^{11}$ s, matter-radiation equality takes place when the energy densities in relativistic (``radiation'') and non relativistic matter (``matter'') are equal. At this moment the evolution of the universe changes from 
radiation dominated to matter dominated. Recombination takes place at around 10$^{12}$ to $10^{13}$ s when the plasma cooled down enough so that electrons and nuclei form neutral atoms. Shortly afterwards the photons decouple from the cosmic plasma, the universe becomes transparent to radiation which today is observed as the cosmic microwave background (CMB). The temperature of the CMB is not perfectly isotropic but there are small deviations from the background isotropy. Moreover, the radiation is linearly polarized (which one would not expect from Thomson scattering of a random, isotropic ensemble of electrons). These CMB anisotropies can be used to test models of the very early universe as will be discussed in the next section. The temperature fluctuations in the CMB are caused by perturbations in the density field. At later times, under the influence of gravity, some of  these fluctuations become gravitationally unstable leading to the first generation of stars which marks the beginning of galaxy and large scale structure formation. This first generation of stars are thought to be important in the reionization of the intergalactic medium (IGM). The ionization state of the IGM can be  determined from the observation of light from distant quasars which are active galactic nuclei with extremely large luminosities.
 The presence of neutral hydrogen in the IGM along the line of sight can be detected by the  observation of Ly$\alpha$ absorption  at wavelengths shorter than $(1+z_*)\lambda_{Ly\alpha}$ in the spectrum of the quasar
 assumed to be  at a redshift $z_*$. Moreover,  $\lambda_{Ly\alpha}=1216$ \AA $\;$is the laboratory wavelength corresponding to the Ly$\alpha$ absorption line. Observations show that this absorption line is absent upto redshifts $z\simeq 6$ indicating that the IGM has been completely reionized below this redshift (c.f., more details in section \ref{sec2}).

The description of physical processes in the early universe such as decoupling and recombination requires to understand the thermodynamics of the early universe. Early on, during the radiation dominated era  the universe is dense and hot. Thermal equilibrium is established by interactions of particles rapid in comparison with  the typical time scale of expansion of the universe.
In thermodynamical equilibrium the number density $n$, the energy density $\rho$ and the pressure $p$ of a gas of particles without strong interactions with $g$ internal degrees of freedom is determined in terms of the distribution function $f(\vec{p})$ in phase space by (cf. e.g., \cite{KT})
\begin{eqnarray}
n&=&\frac{q}{(2\pi)^3}\int f(\vec{p})d^3p\\
\rho&=&\frac{g}{(2\pi)^3}\int E(\vec{p})f(\vec{p})d^3p\\
p&=&\frac{g}{(2\pi)^3}\int\frac{|\vec{p}|}{3E}f(\vec{p})d^3p,
\end{eqnarray}
where $E^2=|\vec{p}|^2+m^2$ with $m$ and $\vec{p}$ the mass and 3-momentum of each particle, respectively.
In kinetic equilibrium the distribution function is given by either the Fermi-Dirac or Bose-Einstein statistics.
Hence 
\begin{eqnarray}
f(\vec{p})=\left[\exp\left(\left(E-\mu\right)/T\right)\pm 1\right]^{-1} 
\end{eqnarray}
where $\mu$ is the chemical potential, +1 refers to the Fermi-Dirac distribution and -1 to Bose-Einstein distribution.
If the particles are in chemical equilibrium and, say, species $i$ interacts with species $j$, $k$, $l$ such that
$i+j\leftrightarrow k+l$ then $\mu_i+\mu_j=\mu_k+\mu_l$. The chemical potential of the photons is set to zero, $\mu_{\gamma}=0$.  Therefore since a particle ($p$) and its antiparticle ($\bar{p}$) annihilate to photons the chemical potential satisfy  $\mu_p=-\mu_{\bar{p}}$. In the relativistic limit $T\gg m$, $T\gg\mu$ the thermodynamical quantities are given by
\begin{eqnarray}
\rho&=&\frac{7}{8}\frac{\pi^2}{30} gT^4\\
n&=&\frac{3}{4}\frac{\zeta(3)}{\pi^2}gT^3
\end{eqnarray}
for fermions and 
\begin{eqnarray}
\rho&=&\frac{\pi^2}{30}gT^4\\
n&=&\frac{\zeta(3)}{\pi^2}gT^3
\end{eqnarray}
for bosons and $\zeta(3)=1.20206$ is the value of the Riemann zeta function. In both cases $p=\frac{1}{3}\rho$ is obtained.
In the non relativistic limit $m\gg T$, for fermions as well as  bosons it is found that 
\begin{eqnarray}
n&=&g\left(\frac{mT}{2\pi}\right)^{\frac{3}{2}}e^{-\frac{m-\mu}{T}}
\label{eq:nnr}
\\
\rho&=&n m\\
p&=&nT,
\end{eqnarray}
where the last relation implies $p\ll\rho$. Obviously, all $N$  particle species in thermodynamical equilibrium contribute
to the total value of the corresponding thermodynamical quantities, so that, e.g., for the energy density,
\begin{eqnarray}
\rho_R=T^4\sum_{i=1}^N\left(\frac{T_i}{T}\right)^4\frac{g_i}{2\pi^2}\int_{x_i}^{\infty}\frac{(u^2-x_i^2)^{\frac{1}{2}}u^2du}{\exp(u-y_i)\pm 1}
\end{eqnarray}
where $u\equiv\frac{E}{T_i}$, $x_i\equiv\frac{m_i}{T_i}$ and $y_i\equiv\frac{\mu_i}{T_i}$. Moreover $T$ is the photon temperature.
The contributions from non relativistic particle species are subleading in comparison to the contributions from the relativistic ones. Therefore to a good approximation the total energy density  is given by
\begin{eqnarray}
\rho_R=\frac{\pi^2}{30}g_*T^4
\label{eq:rhoR}
\end{eqnarray}
and similarly the total pressure $p_R=\frac{1}{3}\rho_R=\frac{\pi^2}{90}g_*T^4$ where $g_*$ counts the total number of relativistic degrees of freedom, $m_i\ll T$,
\begin{eqnarray}
g_*=\sum_{i=bosons}g_i\left(\frac{T_i}{T}\right)^4+\frac{7}{8}\sum_{i=fermions}g_i\left(\frac{T_i}{T}\right)^4.
\end{eqnarray}
This is a function of temperature and hence  of time as massive species are relativistic at high enough temperatures, $T\gg m_i$ but become non relativistic once the temperature drops below their rest mass $T\ll m_i$ at which point they stop contributing.
At $T\ll 1$ MeV the only relativistic species are 3 species of light neutrinos and the photon. As will be discussed in detail below, at that time
photons and neutrinos do not have the same temperature, but rather satisfy
$T_{\nu}=\left(\frac{4}{11}\right)^{\frac{1}{3}}T_{\gamma}$. Thus $g_*(T\ll {\rm MeV})=2\times\frac{7}{8}\times 3\left(\frac{4}{11}\right)^{\frac{4}{3}}+2=3.36$
assuming Dirac neutrinos  which introduces the factor 2 since neutrinos and antineutrinos contribute.
 For temperatures above 1 MeV also positrons and electrons, each contributing two degree of freedom,  are relativistic so that $g_*=2+\frac{7}{8}(2+2+2\times 3)=\frac{43}{4}=10.75$. At temperatures above 300 GeV all species of the standard model are relativistic and $g_*$ is of the order of 100.
 During the radiation dominated epoch $\rho=\rho_R$. Hence using Eqs. (\ref{eq:friedmann}) and (\ref{eq:rhoR}) yields to 
 \begin{eqnarray}
 H^2=1.66 g_*^{\frac{1}{2}}\frac{T^2}{M_P},
\label{eq:HRD}
 \end{eqnarray}
 where $M_P=1.22\times 10^{19}$ GeV is the Planck mass. 
 Primordial nucleosynthesis or big bang nucleosynthesis (BBN)
 predicts the abundances of hydrogen and of light elements such as deuterium, helium-3, helium-4, lithium-7 
which were synthesized at the end of the first three minutes of the universe. These predictions are in good agreement with observations.
BBN provides important constraints on possible deviations from the standard model of cosmology and on new physics Beyond the Standard Model (BSM). One of the key quantities is the ratio of number density of neutrons over the number density of the protons, denoted $\frac{n}{p}$.
In thermodynamical equilibrium it is given by
\begin{eqnarray}
\frac{n}{p}=e^{-\frac{Q}{T}}
\label{eq:np}
\end{eqnarray}
where $Q$ is the difference in the rest masses of neutrons and protons, $Q\equiv m_n-m_p=1.293$ MeV.
Very early on, at very high temperatures, $T\gg 1$MeV, weak interactions are very efficient  so that
$\frac{n}{p}=1$. The neutron-proton interconversion rate  is given by
$\Gamma_{n\leftrightarrow p}\sim G_F^2T^5$ where $G_F=1.166\times 10^{-5}$GeV$^{-2}$  is the Fermi constant. At around a temperature $T_D=0.7$ MeV this conversion rate drops below the expansion rate determined by the Hubble parameter, $H\sim\sqrt{g_*}\,T^2/M_P$. Using Eq. (\ref{eq:np}) it follows that the neutron to proton ratio at freeze-out at around 0.7 MeV is, $\frac{n}{p}=\frac{1}{6}$. After freeze-out this ratio is only changed by the decay of the free neutrons, $n\rightarrow p+e+\nu_e$. The life time of a free neutron is 
$\tau=(885.7\pm 0.8)$ s. After some time $t$ the number density of neutrons is given by $n_n(0)e^{-\frac{t}{\tau}}$. Assuming $t$ to be of the order of $10^2$ s the final value of $\frac{n}{p}$ is given by $\frac{n}{p}=\frac{1}{7}$.
Using this together with the fact that  nearly all neutrons end up in helium-4 yields an estimate of the primordial mass fraction $Y_p$,
\begin{eqnarray}
Y_p=\frac{2\frac{n}{p}}{1+\frac{n}{p}}\simeq 0.25.
\label{eq:Yp}
\end{eqnarray}
The observed value is $Y_p=0.2465\pm 0.0097$. 
The nucleosynthesis chain begins with the formation of deuterium in 
 the process $p(n,\gamma)D$. Because of the low density only 2-body reactions, such as $D(p,\gamma)He^3$, $He^3(D,p)He^4$ are important. The abundances of elements other than helium-4 produced in BBN are in comparison much smaller. Primordial abundances  can be observed
 in metal low regions where the abundances of light elements have not been changed by stellar nucleosynthesis (cf. e.g., \cite{Burles:1999zt}, \cite{tytler}, \cite{Iocco:2008va}).
 
 As described above the Big Bang model fares well with observations. However, there are certain short comings (e.g., \cite{Linde:2005ht, Liddle:2000cg}). These include the flatness problem, that is the observation that at present day the universe is flat.
 For example, Planck 15 found that $|\Omega_K|<0.005$ \cite{planck15-cosmo}.  A simple calculation shows that this requires a severe fine tuning of the initial conditions at the beginning of standard cosmology.
 For the sake of  argument assume the universe to be matter dominated throughout its evolution. Then with Eq. (\ref{eq:oK})
 \begin{eqnarray}
 |\Omega-1|=\frac{|k|}{a^2H^2}\sim t^{\frac{2}{3}}.
 \end{eqnarray}
 Assuming that the total density parameter today $\Omega_0$
 is in the range between $10^{-2}$ and 10 requires that at the time of BBN at around 1s,
 \begin{eqnarray}
 |\Omega-1|_{\rm BBN}=|\Omega-1|_0\left(\frac{t_0}{t_{\rm BBN}}\right)^{-\frac{2}{3}}\Rightarrow
 |\Omega-1|_{\rm BBN}<10^{-11}.
 \end{eqnarray}
 This indicates that a slight change in the initial conditions leads to a completely different universe.
 In addition, the horizon problem encapsulates the fact that domains which have been in causal contact have a limited size. The physical horizon distance is given by
 \begin{eqnarray}
 d_H(t)=a(t)\int_{t_i}^t\frac{dt'}{a(t')}\sim H(t)^{-1}
 \label{eq:dh}
 \end{eqnarray}
 which determines the maximal separation of two points in causal contact. Going back to the time of decoupling   shows that at this moment the universe consisted of many causally disconnected domains. This raises the question of how to explain that the amplitude of the fractional temperature fluctuations of the CMB observed in different directions in the sky is of the order of $\frac{\Delta T}{T}\sim10^{-5}$. These small fluctuations also pose a problem in itself for 
the Friedmann-Robertson-Walker models which are exactly homogeneous and isotropic. However, for large scale structure formation initial  density fluctuations are a necessary ingredient. Therefore the observations of the CMB anisotropies provide a key piece in the puzzle of how galaxies and other structures formed in our universe.

The inflationary paradigm solves these problems of the Big Bang model. Inflation is an era before the beginning of standard cosmology. It is defined as a stage of accelerated expansion of the  universe, $\ddot{a}>0$. De Sitter space-time which is characterized by only a positive cosmological constant and no additional energy density contribution is a typical example of an inflationary solution. In this case $H=\sqrt{\frac{\Lambda}{3}}=const.$ and  effectively, $p=-\rho$. Moreover, in this case the scale factor evolves  from some initial time $t_i$, as
 \begin{eqnarray}
 a(t)=a(t_i)e^{H(t-t_i)}.
 \end{eqnarray}
 The flatness problem is resolved by noticing that in this case the total density parameter is driven towards unity, 
 \begin{eqnarray}
 |\Omega-1|=\frac{|k|}{a^2H^2}\sim |k|e^{-2Ht}\rightarrow 0.
 \end{eqnarray}
 The solution to the horizon problem is provided by the observation that the exponential behaviour
 of the scale factor allows to enlarge one causal domain sufficiently within a finite amount of time.
 Thus the observable universe originates from one patch in causal contact in the very early universe.
 The origin of the density fluctuations are assumed to be quantum fluctuations of a scalar field.
 One of the simplest and most studied models of inflation is driven by the potential energy $V(\phi)$ of such a scalar field $\phi$ \cite{Linde:2005ht}. 
 Assuming that the only matter in the universe is given by this scalar field implies that the energy density and pressure are given by,
 \begin{eqnarray}
 \rho&=&\frac{\dot{\phi}^2}{2}+V(\phi)\\
 p&=&\frac{\dot{\phi}^2}{2}-V(\phi).
 \end{eqnarray}
 Going back to Eq. (\ref{eq:ddota}) it can be checked that $\ddot{a}>0$ since $\rho+3p<0$
 in this case.
 The evolution of the scalar field is determined by the Klein-Gordon equation,
 \begin{eqnarray}
 \ddot{\phi}+3H\dot{\phi}=-\frac{dV}{d\phi}.
 \label{eq:KG}
 \end{eqnarray}
 The Friedmann equation takes the form
 \begin{eqnarray}
 H^2=\frac{8\pi}{3M_P^2}\left(\frac{\dot{\phi}^2}{2}+V(\phi)\right).
 \end{eqnarray}
 One important class of inflationary models uses the slow roll approximation which consists
 in neglecting the $\ddot{\phi}$ term in Eq. (\ref{eq:KG}) as well as the kinetic energy term in the Friedmann
 equation, so that
 \begin{eqnarray}
 3H\dot{\phi}&=&-\frac{dV}{d\phi}\\
 H^2&=&\frac{8\pi}{3M_P^2}V(\phi).
 \end{eqnarray}
 This results in the expansion  of the universe being driven by the potential energy of the 
 scalar field which is also called the inflaton. There is no clear candidate from particle physics
 as to what could be the inflaton. It remains one of the challenges to connect inflation with fundamental particle physics.
 The potential is typically chosen to be an even power of $\phi$ and the picture is that the scalar field is slowly rolling down its potential.
 The possibility for slow roll inflation is determined by the shape of the potential which is encoded in the slow roll parameters (cf. e.g. \cite{Lyth:2009zz}),
 \begin{eqnarray}
 \epsilon(\phi)&=&\frac{M_P^2}{16\pi}\left(\frac{V'}{V}\right)^2\\
 \eta(\phi)&=&\frac{M_P^2}{8\pi}\frac{V''}{V}.
 \label{eq:slowroll}
 \end{eqnarray}
 The conditions for slow roll inflation are $\epsilon\ll 1$ and $|\eta|\ll 1$.
 The field value at the end of inflation is determined by the condition that 
  the modulus of at least one of the slow parameters reaches unity.
 The duration and hence the amount of inflation is measured by the number of e-folds,
 \begin{eqnarray}
 N(t)=\ln\frac{a(t_{fin})}{a(t)}
 \end{eqnarray}
 where $t_{fin}$ denotes the end of inflation. In slow roll inflation 
 this can be expressed as a function of $\phi$,
 \begin{eqnarray}
 N(\phi)=\frac{8\pi}{M_P^2}\int_{\phi}^{\phi_{fin}}d\phi\frac{V}{V'}.
 \end{eqnarray}
 The requirement for successful resolution of the problems with standard cosmology puts a lower bound on the
 total amount of e-folds which is typically in the range between  55 and 65.
 In the inflationary paradigm the temperature fluctuations and polarization of the CMB  as well as the seeds necessary for large scale structure formation have their origin in the quantum fluctuations of the inflaton.
 During inflation the physical horizon size determined by $H^{-1}$ (cf. Eq. (\ref{eq:dh})) is approximately a constant.
 This means that physical wavelengths which were within the horizon at some time are stretched beyond the horizon at some later time (recall that all physical scales are comoving scales multiplied by the scale factor). The horizon crossing takes place at $\lambda=1/(aH)$ where $\lambda$ denotes the comoving wavelength.
 Once a perturbation is outside the horizon its amplitude freezes and  it becomes a classical perturbation.
 The spectrum of fluctuations can be calculated by quantizing the inflaton on the homogeneous background. Quantization of a scalar field on a de Sitter background is a well studied problem. Of course, slow roll inflation is not exactly a de Sitter background but the spectrum can still be calculated approximately. It is interesting to note that in the case of power law inflation for which the scale factor evolves as $a\sim t^p$ and the potential is an exponential potential the spectrum can be found exactly (for a detailed discussion of this case see, e.g, \cite{PU}).
 
 The scalar field can be separated into a homogeneous part $\phi(t)$ and a perturbative part $\delta\phi(t,\vec{x})$ such that
 $\phi(t,\vec{x})=\phi(t)+\delta\phi(t,\vec{x})$. 
 Considering the simplest case of a massless scalar field then the perturbations satisfy in Fourier space for each comoving wave number $\vec{k}$  the mode equation,
 \begin{eqnarray}
 \ddot{\delta\phi}_{\vec{k}}+3H\dot{\delta\phi}_{\vec{k}}+\left(\frac{k}{a}\right)^2\delta\phi_{\vec{k}}=0.
 \label{eq:deltaphi}
 \end{eqnarray}
 Quantizing $\delta\phi$ leads to a 2-point function $\langle\delta\phi_{\vec{k}}\delta\phi_{\vec{k}'}\rangle=\left(\frac{H}{2\pi}\right)^2 \delta_{\vec{k}\vec{k'}}$.
 Therefore on super horizon scales classical fluctuations in the scalar field with an amplitude $|\delta\phi|\simeq\frac{H}{2\pi}$  are induced. This in turn induces curvature perturbations which are imprinted in the CMB. This will be discussed in more detail in the next section.
 The phases of each wave are random. At each point in space the sum of all waves is described by a Brownian motion in all directions implying Gaussian perturbations.

 \section{The inhomogeneous universe}
 \label{sec3}

 The cosmic microwave background provides us with a unique window to the physics of the early universe.
 To understand its formation it is important to understand the thermal and ionization history of the cosmic plasma. As the universe expands its temperature $T$ corresponding to the photon temperature cools down as $T\propto a^{-1}$. Very early on, deep inside the radiation dominated era, temperatures are high enough so that all particles in the cosmic plasma are strongly coupled by interactions. In the best fit $\Lambda$CDM model  
 the initial conditions for the numerical evolution of  the perturbations which are imprinted in the CMB as temperature anisotropies and polarization are set after neutrino decoupling at around 1 MeV. At this time matter in the universe consists of a strongly coupled photon-baryon fluid and cold dark matter. To complete the model which best fits the data a cosmological constant $\Lambda$ has to be added. However, dynamically $\Lambda$ does not play a role until very close to the present epoch. Dark matter only interacts gravitationally and its presence is required to provide the gravitational potential field to explain the observed curvature fluctuations.
 
 Once the temperature of the primordial plasma is low enough electrons and nuclei, mostly protons, recombine to form neutral atoms, mostly hydrogen. Defining the epoch of recombination by requiring that the ionization fraction is 0.1 determines the corresponding redshift to be $z=1360$ and the temperature is of the order of 4000K. This is 
much lower than what would be expected if it just depended on the ionization energy of hydrogen 13.6 eV which corresponds to about 160000 K. This is due to subtleties in the recombination process involving two-photon decay processes (cf., e.g., \cite{Challinor:2009tp}). The epoch of decoupling is described to be the moment after which (most) photons will not scatter again. Though this is not true for all photons since there is a residual ionization fraction of about $10^{-4}$.
To be more precise  photon decoupling is defined to be  the epoch when the time between scatterings equals the age of the universe.
 This defines the surface of last scattering and is the origin of what is observed today as the cosmic microwave background (CMB).  As already mentioned the CMB is remarkably homogeneous and isotropic. However, there are small temperature fluctuations $\Delta T/T\sim 10^{-5}$ and it is linearly polarized. 
 It is precisely these temperature anisotropies and polarization which constrain the physics of the very early universe
 and its evolution. As will be discussed later on in more detail at around a redshift $z=10$ reionization takes place which generates additional CMB anisotropies and polarization.
 
 Temperature fluctuations $\Theta(\hat{\mathbf n})=\delta T/T$ in the direction $\hat{\mathbf n}$ on the sky are expanded in spherical harmonics  $Y_{\ell m}(\hat{\mathbf n})$ such that (cf. e.g. \cite{HW})
 \begin{eqnarray}
 \Theta(\eta,{\bm x},\hat{\bm n})=\int\frac{d^3k}{(2\pi)^3}\sum_{\ell}\sum_{m=-2}^2\Theta_{\ell}^{(m)}G_{\ell}^m
 \label{theta}
 \end{eqnarray}
 which in general includes contributions from the scalar ($m=0$), vector ($m=\pm 1$) and tensor ($m=\pm 2$) modes which are uncoupled at linear order. 
 These modes describe the linear perturbations of the metric as well as the energy momentum tensor in Fourier space. They will be discussed below in more detail.
 Moreover, 
 \begin{eqnarray}
 G_{\ell}^m=(-i)^{\ell}\sqrt{\frac{4\pi}{2\ell+1}}Y^m_{\ell}(\hat{\bm n})e^{i{\bm k}\cdot{\bm x}}.
 \end{eqnarray}
The two-point function is determined by the corresponding angular power spectrum $C_{\ell}^{XY}$ such that
\cite{HW}
\begin{eqnarray}
(2\ell+1)C_{\ell}^{XY}=\frac{2}{\pi}\int \frac{dk}{k}\sum_{m=-2}^{2}k^3\langle X_{\ell}^{(m)*}(\eta_0,k)
Y_{\ell}^{(m)}(\eta_0,k)\rangle
\end{eqnarray}
where $X$ and $Y$ denote $\Theta$, $E$, and $B$. 
For completeness, also the polarization modes $E$ and $B$ are included here.
The polarization of the CMB will be discussed in more detail below.

 The cosmic microwave background has a nearly perfect Planck spectrum at a temperature
$T_0=2.7255\pm 0.0006$ K at 1-$\sigma$ \cite{Fixsen:2009ug} which corresponds to the monopole, $\ell=0$. 
The dipole corresponding to the multipole $\ell=1$ has the largest amplitude of the temperature fluctuations at
$3.372\pm 0.014$ mK (95\ CL) \cite{firas3}.
It is due to the motion of the solar system with respect to the CMB.
An observer moving with a velocity $\beta=\frac{v}{c}$ relative to an isotropic Planck radiation field of temperature $T_0$ measures a Doppler shifted temperature pattern,
\begin{eqnarray}
T(\theta)=\frac{T_0(1-\beta^2)^{\frac{1}{2}}}{1-\beta\cos\theta}\simeq T_0\left(1+\beta\cos\theta+\frac{\beta^2}{2}\cos(2\theta)+\frac{\beta^2}{2}\cos 2\theta+{\cal O}(\beta^3)\right)
\end{eqnarray}
 observing at every point in the sky a black body spectrum with temperature $T(\theta)$.
 The observed Doppler shift implies that the barycenter of the solar system is moving with a velocity $371\pm 1$ km s$^{-1}$ relative to the CMB rest frame in the direction determined by the galactic coordinates $(l,b)=(264.14^º\pm 0.15^º, 48.26^º\pm 0.15^º)$ which is almost orthogonal to the direction of the Galactic center \cite{firas3}. 
 Since the dipole is a frame-dependent quantity it is possible to define an "absolute rest frame" in which  the CMB
 dipole is zero.
 Higher order order multipoles, $\ell\geq 2$ carry the information about fluctuations in the matter density and velocity fields as well as the  gravitational field from before last scattering of the present day CMB photons upto today.
 There are quite a large number of  observations of the CMB: ground based with radio telescopes, detectors mounted on balloons and satellites. To mention a few examples, the Atacama Cosmology Telescope (ACT) is a 6m radio telescope situated in the Atacama Plateau in Chile. It observes the CMB over a large area of the sky at three frequency channels, 148 GHz, 218 GHz and 277 GHz. Observations of the CMB anisotropies are for multipoles $540<\ell<9500$ \cite{act}. Another example is the South Pole Telescope (SPT) located at the south pole and observing similarly the temperature anisotropies for multipoles $2000<\ell<9400$ at three frequency bands, namely, 95 GHz , 150 GHz and 220 GHz \cite{spt}.
 Boomerang was the first balloon experiment to observe the CMB. There were two flights (1998, 2003)
 launched from McMurdo Station on a circular path around Antarctica lasting about 10 days \cite{boomerang}. The angular power spectrum of the CMB anisotropies was obtained for $75<\ell<1025$ in four 150 GHz channels.
 The first satellite to observe the CMB was COBE (Cosmic Background explorer) which was launched by NASA in 1989. 
Its measurement of the absolute spectrum of the CMB revealed a nearly perfect black body spectrum. The tiny deviations from the Planck spectrum  observed by the COBE/FIRAS instrument constitute the first observational constraints on spectral distortion of the CMB. 
There are proposals for  future space missions, such as PIXIE (Primordial Inflation Explorer) \cite{pixie}, to measure and tighten the constraints on spectral distortions of the CMB together with precise polarization measurements (B-mode).
COBE observed the CMB temperature anisotropies at an effective angular resolution of 10 degrees in three 
frequency channels (31.5 GHz, 53 GHz and 90 GHz) \cite{COBE}.
After COBE the next satellite experiment to observe the CMB temperature anisotropies as well as polarization was WMAP (Wilkinson Microwave Anisotropy Probe) operated by NASA between 2001 and 2010 observing in 5 frequency channels (23,33,41,61 and 94 GHz). Angular power spectra cover a multipole range of $2\leq\ell<1000$ \cite{wmap}.
 The Planck mission was on board a satellite operated by ESA. It took data between 2009 and 2013. 
 It observed the CMB temperature anisotropies and polarization. The temperature anisotropy angular power spectrum covers the range $2\leq\ell<2500$ observed in separate frequency channels covering the range 30-857 GHz \cite{planck}.
 
 The observed CMB temperature perturbations are the result of perturbations of the perfectly isotropic and spatially homogeneous Friedman-Robertson-Walker background and its matter distribution. The scales on which the CMB is observed are sufficiently large so that perturbations are still in the linear regime. 
At linear order there are three types of perturbations of an FRW background, namely, scalar, vector and tensor perturbations depending on their behaviour under general coordinate transformations.
Starting with a flat FRW background determined by the line element,
\begin{eqnarray}
ds^2=a^2(\eta)(d\eta^2-\delta_{ij}dx^idx^j)
\end{eqnarray}
 the most general linear perturbation of the metric is given by 
 \begin{eqnarray}
 ds^2=a^2(\eta)\left[\left(1-2A\right)d\eta^2+2B_id\eta dx^i-\left[\left(1+2D\right)\delta_{ij}+2E_{ij}\right)dx^idx^j\right],
\label{eq:linpert}
\end{eqnarray}
 where the metric perturbation variables $A$, $B_i$, $D$ and $E_{ij}$ are all functions of space and time and the Einstein summation convention is used which corresponds to summing over repeated indices. Moreover, latin indices take values between 1 and 3.
 These functions are expanded in scalar, vector and tensor harmonics. 
In the case of scalar and vector perturbations there is an inherent gauge freedom associated with the metric perturbation variables.
 This is related to the fact that when perturbing the metric $g_{\mu\nu}\rightarrow g_{\mu\nu}+\delta g_{\mu\nu}$
 this can be done using different ways of defining a coordinate system or in other words, slicings, corresponding to the choice of equal time hypersurfaces. Therefore, two perturbations of the same FRW background, with metric tensor, say ${\cal G}_{\mu\nu}$ and ${\cal H}_{\mu\nu}$, are related by a transformation of the space-time coordinates, $x^{\mu}\rightarrow \tilde{x}^{\mu}$ and the corresponding transformation of the metric tensors, ${\cal G}_{\mu\nu}(x^{\alpha})\rightarrow {\cal H}_{\mu\nu}(\tilde{x}^{\alpha})$.  Since these are perturbations of the same background space-time the change in the perturbed metric tensor has to be calculated at the same coordinate value, 
that is $\Delta{\cal T}_{\mu\nu}(x^{\alpha})\equiv{\cal H}_{\mu\nu}(x^{\alpha})-{\cal G}_{\mu\nu}(x^{\alpha})$. 
Calculating this to first order in the perturbations determines the gauge transformations of the  metric perturbation variables (cf. Eq. (\ref{eq:linpert})) as well as the  corresponding linear perturbations of the energy-momentum tensor.
 Naturally, the physics must not depend on a gauge choice so all relevant quantities have to be gauge-invariant.
Whereas for scalar modes there exists a number of gauge choices, for vector modes there exist only two gauge choices and the tensor modes are described by only gauge-invariant quantities. In the following we focus on the scalar mode and 
choose the so called conformal Newtonian gauge defined below, in Eq. (\ref{eq:conNewgauge}), as well as the gauge-invariant formulation.
 The perturbation equations are then derived by calculating the first order perturbation of Einstein's equations. For details see, e.g, \cite{KS} or \cite{Lyth:2009zz},
 \cite{durrer}, \cite{PU}, \cite{Weinberg:2008zzc}.
 
 Long before their decoupling photons are tightly coupled to the rest of the cosmic plasma by Thomson scattering. This implies that they are in thermal equilibrium and have a Planck distribution. However, as the universe cools down, Thomson scattering becomes less efficient and photons fall out of equilibrium perturbing the Planck  distribution.
 In order to calculate the CMB temperature anisotropies and polarization of the CMB one has to follow the evolution of the photon phase-space distribution in the perturbed FRW background $f(\eta,{\mathbf x},{\mathbf n},q)$,
 given by 
 \begin{eqnarray}
 f(\eta,{\mathbf x},{\mathbf n},q)=f(q)+\delta f(\eta,{\mathbf x},{\mathbf n},q),
 \end{eqnarray} 
 where $q=a(\eta)p(\eta,{\mathbf x})$ is the comoving photon energy and $f(q)$ the black body distribution,
 \begin{eqnarray}
 f(q)=\frac{1}{e^{\frac{q}{T_0}}-1}.
 \end{eqnarray}
 Moreover ${\mathbf n}$ points along the direction of propagation of the photons.
 The distribution function is determined by the Boltzmann equation including a collision term $C[f]$ (cf., e.g., \cite{Lyth:2009zz})
 \begin{eqnarray}
 \frac{\partial f}{\partial\eta}+\frac{\partial f}{\partial x^i}\frac{dx^i}{d\eta}+\frac{\partial f}{\partial q}
 \frac{dq}{d\eta}+\frac{\partial f}{\partial n^i}\frac{dn^i}{d\eta}=\frac{df}{d\eta}\equiv C[f].
 \end{eqnarray}
Defining the brightness function $\Theta\equiv\frac{\delta T}{T}$ yielding 
 \begin{eqnarray}
 f(\eta,{\mathbf x},{\mathbf n},q)=\frac{1}{e^{\frac{q}{T_0(1+\Theta)}}-1}
 \end{eqnarray}
 results at  first order in 
 \begin{eqnarray}
 \delta f(\eta,{\mathbf x},{\mathbf n},q)=-q\frac{df(q)}{dq}\Theta(\eta,{\mathbf x},{\mathbf n}).
 \end{eqnarray}
 The photon trajectory in the perturbed FRW background can be expressed at first order in terms of the 
 functions characterizing the linear perturbations of the metric (cf., Eq. (\ref{eq:linpert})). Since at linear order the perturbation equations for the scalar, vector and tensor modes decouple so do the corresponding Boltzmann equations. 
 The scalar mode is the most important one to establish the (minimal) $\Lambda$CDM model. Thus, as way of example, we will consider the Boltzmann equation derived for the scalar mode perturbation of a flat FRW background.
 The scalar mode perturbation of the metric Eq.  (\ref{eq:linpert}) in the conformal Newtonian gauge is given by
 \begin{eqnarray}
 ds^2=a^2(\eta)\left[-\left(1+2\Psi\right)d\eta^2+\left(1-2\Phi\right)\delta_{ij}dx^idx^j\right]
 \label{eq:conNewgauge}
 \end{eqnarray}
 where $\Phi$ and $\Psi$ are called gravitational potentials. This is related to the fact that during the matter dominated era $\Phi=\Psi$ and Newtonian gravity applies on scales well within the horizon with
 $\Phi$ being the Newtonian gravitational potential.
 For the scalar mode perturbation, the Boltzmann equation for the brightness perturbation at linear order in Fourier space, $\Theta(\eta,{\mathbf k},{\mathbf n})$, is given by
 \begin{eqnarray}
 \frac{\partial\Theta}{\partial\eta}+ik\mu\Theta-\frac{\partial\Phi}{\partial\eta}+ik\mu\Psi=C[\Theta]
 \end{eqnarray}
 where $\mu=\hat{\mathbf k}\cdot{\mathbf  n}$. Similarly, a Boltzmann equation for the vector as well as the tensor mode can be derived (for details, see e.g., \cite{KS} or \cite{Lyth:2009zz},
 \cite{durrer}, \cite{PU}).
 The next step is to expand the brightness perturbation $\Theta(\eta,{\mathbf x},{\mathbf n})$ in terms of spherical harmonics  (cf. equation (\ref{theta})). 
 This leads to the Boltzmann hierarchy which determines the evolution of the multipole components $\Theta_{\ell}^{(m)}(\eta,{\mathbf k})$. For the scalar mode  the first three multipoles determine  the  energy density perturbation, $\Theta_0^{(0)}=\frac{\delta_{\gamma}}{4}$, the velocity, $\Theta_1^{(0)}=\frac{V_{\gamma}}{3}$, and the anisotropic stress of the photon fluid, $\Theta_2^{(0)}=\frac{\pi_{\gamma}}{12}$. The latter one encodes one of the contributions to the deviation from a perfect fluid.
 There exist a variety of numerical programs to solve the Boltzmann hierarchy and calculate the CMB anisotropies.
 These are all open source. The first one was the {\tt COSMOS} program \cite{Bertschinger:1995er}
 followed by {\tt CMBFAST} \cite{Seljak:1996is}, {\tt CAMB} \cite{Lewis:1999bs}, {\tt CMBEASY} \cite{Doran:2003sy}
 and {\tt CLASS} \cite{class1,class2,class3,class4} to mention just a few. A good reference to find links to these and other programs as well as to published data of many CMB observations is the NASA Legacy Archive for Microwave Background Data Analysis {\tt LAMBDA} at {\tt http://lambda.gsfc.nasa.gov/}.
 
 The minimal model used to fit the observations of the CMB is the 6-parameter $\Lambda$CDM model in which case 
 initial conditions for the numerical solutions are set after standard model (SM) neutrino decoupling, thus at $T<1$ MeV. At this moment, well within the radiation dominated era, in the universe the matter is constituted by the already decoupled SM neutrinos, cold dark matter (CDM) and  the tightly coupled baryon-electron-photon fluid.  
In the latter matter component the Coulomb interaction between electrons and baryons, that is nuclei, mostly protons (75\%) and helium-4 nuclei (<25\%) cause locally equal number densities  of electrons and baryons.
Photons are tightly coupled to the electron-baryon fluid via the (non relativistic) Thomson scattering off the (already non-relativistic) electrons and nuclei.
 The minimal model is based on the scalar mode with adiabatic initial conditions which will be discussed below. 
 Extensions of the minimal $\Lambda$CDM model include a tensor mode which  would be generated if the relevant cosmological perturbations are generated during inflation. The tensor mode also has a particular signature in the polarization of the CMB as it generates the B-mode of polarization (see below) which is why the observation of a primordial B-mode is considered to be a strong indication that inflation took place in the early universe.
 Other extensions include  massive neutrinos (cf., e.g., \cite{Lesgourgues:2006nd}) and primordial magnetic fields, e.g., \cite{Mack:2001gc,Shaw:2009nf,Kunze:2011bp}.
Primordial magnetic fields source all three types of linear perturbation modes. In particular, whereas vector modes decay in the standard $\Lambda$CDM model they do get sourced by a primordial magnetic field present before decoupling.

Initial conditions for the cosmological perturbation equations are set well outside the horizon deep inside the radiation dominated era. For scalar modes adiabatic initial conditions and isocurvature initial conditions constitute two different classes of initial conditions. Taking as an example the radiation-baryon fluid then its specific entropy of radiation normalized to the baryon number $n_b$ is given by $S_{\gamma b}=\frac{S_{\gamma}}{n_b}$. Hence the first order fractional  perturbation is given by (cf. e.g. \cite{2011pcmb.book.....N})
\begin{eqnarray}
\frac{\delta S_{\gamma b}}{S_{\gamma b}}=\frac{\delta S_{\gamma}}{S_{\gamma}}-\frac{\delta n_b}{n_b}=3\frac{\delta T_{\gamma}}{T_{\gamma}}-\frac{\delta n_b}{n_b}=\frac{\delta\rho_{\gamma}}{\rho_{\gamma}+p_{\gamma}}-\frac{\delta\rho_b}{\rho_b}=\frac{3}{4}\delta_{\gamma}-\delta_b.
\label{eq:deltaS}
\end{eqnarray}
In this case imposing adiabaticity implies 
\begin{eqnarray}
\frac{\delta S_{\gamma b}}{S_{\gamma b}}=0\Rightarrow \frac{3}{4}\delta_{\gamma}=\delta_b.
\end{eqnarray}
Eq. (\ref{eq:deltaS}) can be generalized to any other fluid component with equation of state, $p_j=w_j\rho_j$, where $w_j$ is a  constant, so that in general,
\begin{eqnarray}
\frac{\delta S_{\gamma j}}{S_{\gamma j}}=\frac{\delta_{\gamma}}{w_{\gamma} +1}-\frac{\delta_j}{1+w_j}.
\end{eqnarray}
Thus for the $\Lambda$CDM model adiabatic initial conditions are determined by
\begin{eqnarray}
\frac{1}{4}\delta_{\gamma}=\frac{1}{4}\delta_{\nu}=\frac{1}{3}\delta_b=\frac{1}{3}\delta_{cdm}.
\end{eqnarray}
The amplitude of adiabatic perturbations are characterized by  a gauge-invariant quantity, namely, the curvature perturbation on a uniform density hypersurface $\zeta$. 
In the gauge invariant formalism it is given by (cf., e.g., \cite{durrer})
\begin{eqnarray}
-\zeta=-\Phi+{\cal H}k^{-1}V
\end{eqnarray}
which for the standard $\Lambda$CDM yields to
\begin{eqnarray}
\zeta=\frac{\Delta_{\gamma}}{4},
\end{eqnarray}
were $\Delta_{\gamma}$ is the gauge-invariant photon energy density contrast.
On large scales,  this is related to the  curvature perturbation on a comoving slicing by ${\cal R}\simeq -\zeta$ (cf., e.g., \cite{PU}).

The minimal, 6-parameter  $\Lambda$CDM model includes only the scalar, adiabatic mode to source the density fluctuations necessary to explain the observed CMB temperature anisotropies and polarization. 
One of the quests of cosmology is to determine the origin of the corresponding primordial curvature fluctuations.
A promising candidate is single field inflation as it naturally generates adiabatic initial conditions. At the end of inflation 
the inflaton decays and the SM particles are created. Since their overall ratios are fixed, this leads to
$\delta(n_A/n_B)=0$ implying $\delta S_{A,B}/S_{A,B}=0$ \cite{Langlois:2010xc, Weinberg:2004kr}.
The comoving curvature perturbation from single field inflation is give by
\begin{eqnarray}
{\cal R}=-H\frac{\delta\rho}{\dot{\rho}}\Rightarrow{\cal R}=-H\frac{\delta\phi}{\dot{\phi}}.
\end{eqnarray}
Using the two point function of the inflaton fluctuations (cf. after Eq. (\ref{eq:deltaphi})) the  two-point function $\langle{\cal R}_{\mathbf k}{\cal R}_{\mathbf k'}\rangle=\frac{2\pi^2}{k^3}{\cal P}_{{\cal R}}\delta_{{\mathbf k},{\mathbf k'}}$ is then determined by the power spectrum 
\begin{eqnarray}
{\cal P}_{{\cal R}}(k)=\frac{1}{4\pi}\left.\left(\frac{H^2}{\dot{\phi}}\right)^2\right|_k
\end{eqnarray}
which in slow roll inflation can be expressed as \cite{Liddle:2000cg}
\begin{eqnarray}
{\cal P}_{{\cal R}}(k)=\frac{1}{24\pi^2M_P^4}\left.\frac{V}{\epsilon}\right|_{k=aH}
\end{eqnarray}
where $\epsilon$ is one of the slow roll parameters defined in Eq. (\ref{eq:slowroll})
calculated at horizon crossing.
An effective spectral index can be defined as 
\begin{eqnarray}
n-1\equiv\frac{d\ln{\cal P}_{{\cal R}}(k)}{d\ln k}
\end{eqnarray}
which can be expressed in terms of the slow roll parameters
as 
\begin{eqnarray}
n(k)-1=-6\epsilon+2\eta.
\end{eqnarray}
In addition often a running of the spectral index is considered,
\begin{eqnarray}
\frac{dn}{d\ln k}=-16\epsilon\eta+24\epsilon^2+2\xi
\end{eqnarray}
where a third slow roll parameter $\xi$ is defined by $\xi\equiv\frac{M_P^4}{64\pi^2}\frac{V'V'''}{V^2}$.
For example, Planck 15 constrains this scale dependence of the spectral index as \cite{planck15-cosmo}
\begin{eqnarray}
\frac{dn}{d\ln k}&=&-0.0084\pm0.0082, \hspace{2cm} Planck \,TT+lowP\\
\frac{dn}{d\ln k}&=&-0.0057\pm 0.0071, \hspace{2cm} Planck\, TT, TE, EE+lowP
\end{eqnarray}
where $TT$ denotes the Planck data determining the temperature autocorrelation function, $TE$ the Planck temperature and E-mode polarization cross correlation data, $EE$ the Planck E-mode auto correlation data and $lowP$ the   low $\ell$ E-mode polarization data. 

Isocurvature initial conditions have the defining property of a vanishing total curvature perturbation, ${\cal R}=0$.
They do, however, imply entropy perturbations between the photons and other types of matter in the universe
as discussed before.
In general, these can be generated during multi-field inflation since single-field inflation only excites adiabatic modes. 
Hence isocurvature initial conditions are determined by  
\begin{eqnarray}
S_j\equiv\frac{\delta S_{\gamma j}}{S_{\gamma j}}\neq 0
\end{eqnarray} 
where the index $j$ indicates the type of matter.
Each of these defines an independent isocurvature mode. In the minimal extension of the $\Lambda$CDM model  isocurvature modes are induced by fluctuations of the baryon number w.r.t to the photon number density, namely the baryon isocurvature mode (${\cal R}=0$, $S_{cdm}=0=S_{\nu}$, $S_b\neq 0$), as well as the CDM isocurvature mode (${\cal R}=0$, $S_b=0=S_{cdm}$, $S_{cdm}\neq 0$). In principle there are also two additional isocurvature modes for the neutrinos.
One is the neutrino density isocurvature mode in which case the initial conditions are determined by 
${\cal R}=0$, $S_b=0=S_{cdm}$, $S_{\nu}\neq 0$. It can be excited similarly to the baryon or CDM isocurvature modes.
The other is the neutrino velocity mode taking into account the evolution of the neutrino anisotropic stress. However, so far there is no physical mechanism to excite this mode (cf., e.g.,\cite{planck15-inflation}, \cite{LMMP}).

The resulting angular power spectrum of only isocurvature modes is  not compatible with observations.
Though, it is possible to have a relatively small contribution from isocurvature modes with the total power spectrum clearly dominated by the adiabatic mode. In general, the different modes can have non-vanishing correlations.
Thus there are two parameters to characterize the isocurvature mode contribution. Firstly, there is the primordial isocurvature fraction which in general depends on scale in the parametrization used in \cite{planck15-inflation},
\begin{eqnarray}
\beta_{iso}(k)=\frac{{\cal P_{II}}(k)}{{\cal P}_{\cal R}(k)+{\cal P}_{\cal II}(k)}
\end{eqnarray}
where ${\cal I}$ denotes the isocurvature mode. Secondly, there is the correlation between the adiabatic mode (${\cal R}$) and the 
isocurvature mode $({\cal I})$ which is encoded in the scale independent correlation fraction,
\begin{eqnarray}
\cos\Delta_{ab}=\frac{{\cal P}_{ab}}{({\cal P}_{aa}{\cal P}_{bb})^{\frac{1}{2}}}
\end{eqnarray}
where $a,b = {\cal R}, {\cal I}$.

For example, the Planck collaboration finds from the  analysis of the 2015 temperature data plus polarization at low $\ell$ for the adiabatic mode plus a CDM isocurvature mode that the 95\% CL upper bounds on the scale dependent fractional primordial contribution of isocurvature modes are \cite{planck15-inflation}
\begin{eqnarray}
100\beta_{iso,k=0.002 {\rm Mpc}^{-1}}=4.1, 
\;\;\;
100\beta_{iso,k=0.050 {\rm Mpc}^{-1}}=35.4,
\;\;\;
100\beta_{iso,k=0.100 {\rm Mpc}^{-1}}=56.9
\end{eqnarray}
and the scale independent primordial correlation fraction, $\cos\Delta_{{\cal R}{\cal I}}$ is in the intervall [-30:20].

\begin{figure}
\centering{\includegraphics[width=0.32\linewidth]{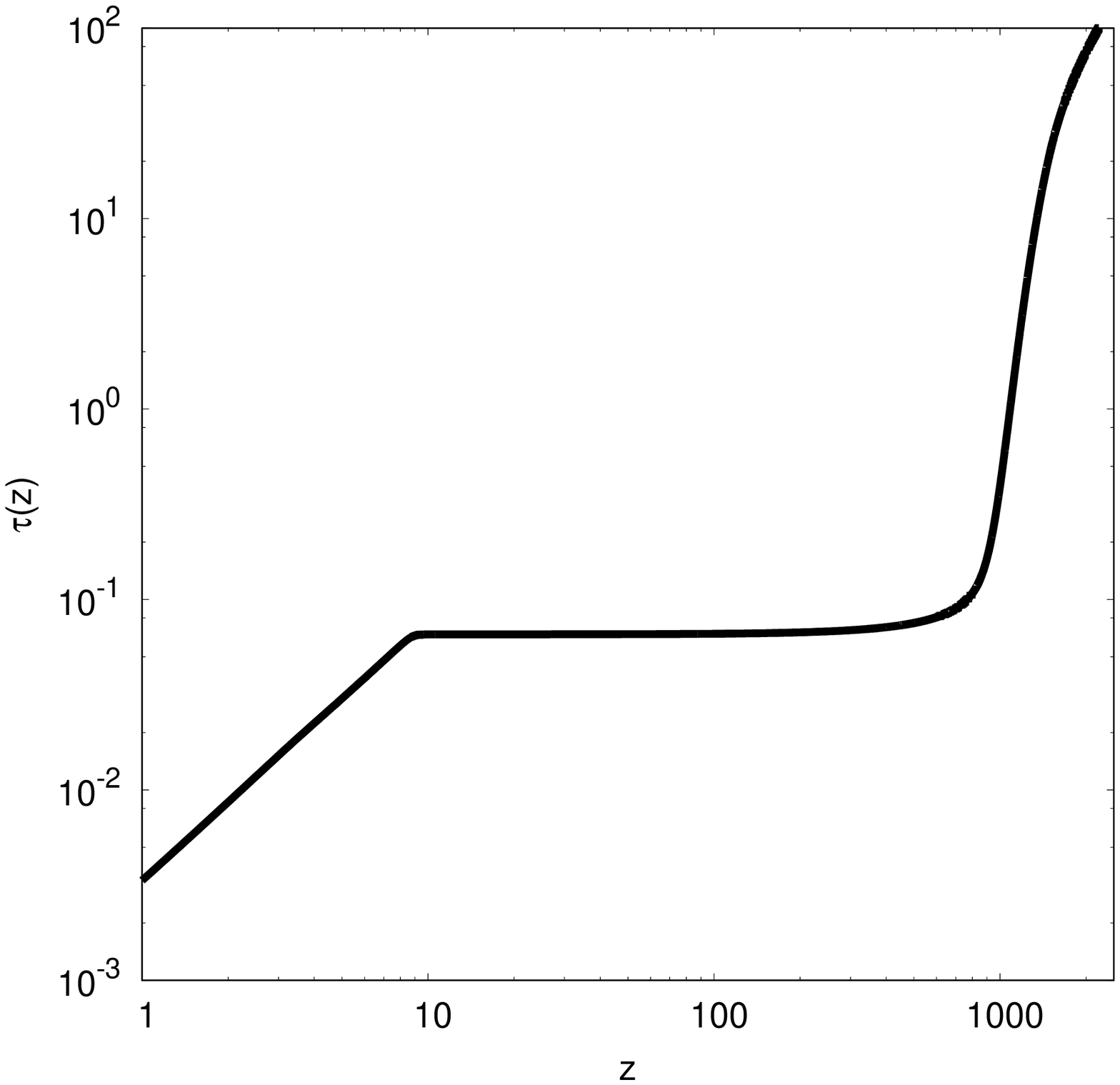}\hspace{0.01cm}
\includegraphics[width=0.32\linewidth]{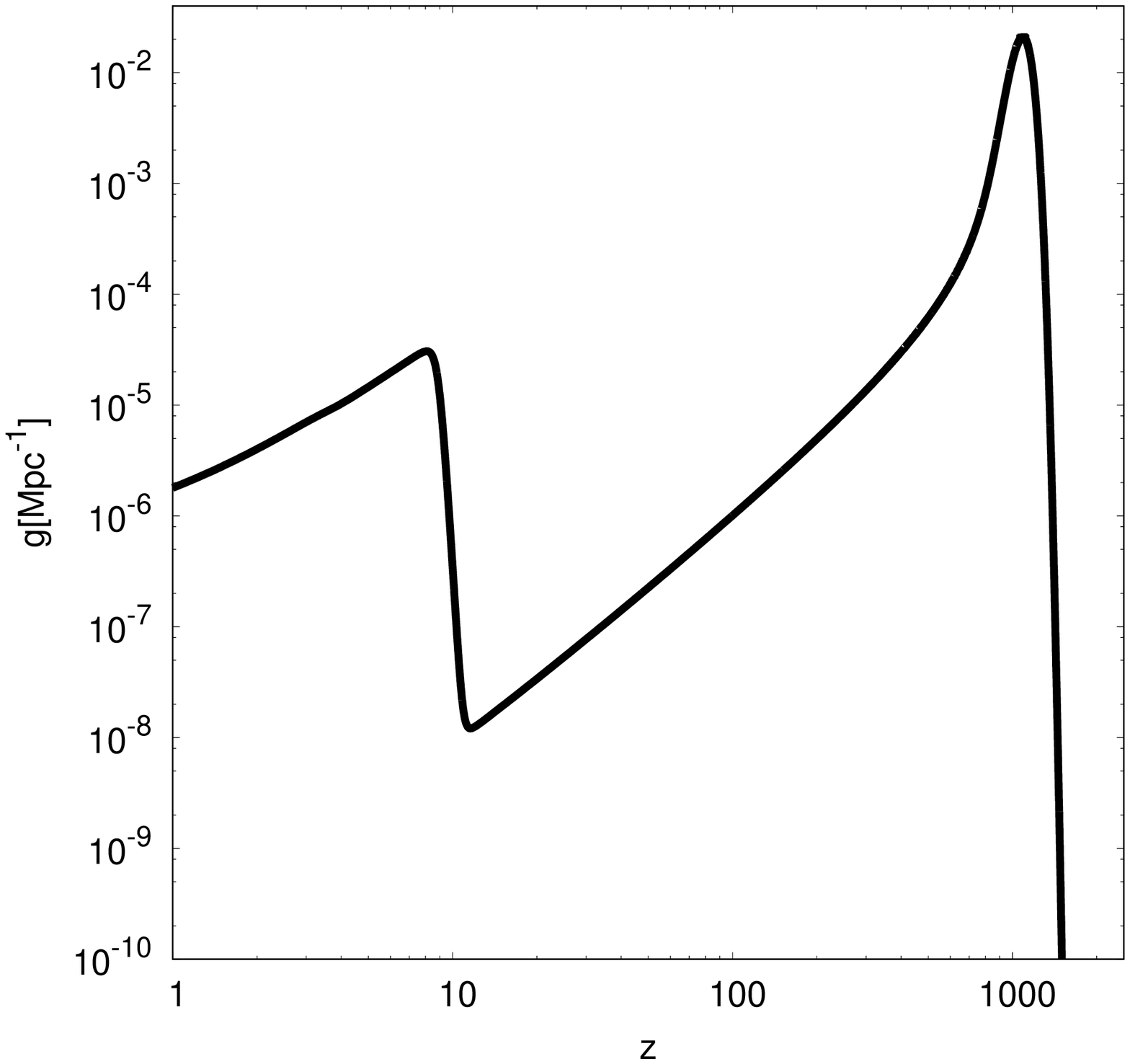}
\hspace{0.01cm}
\includegraphics[width=0.32\linewidth]{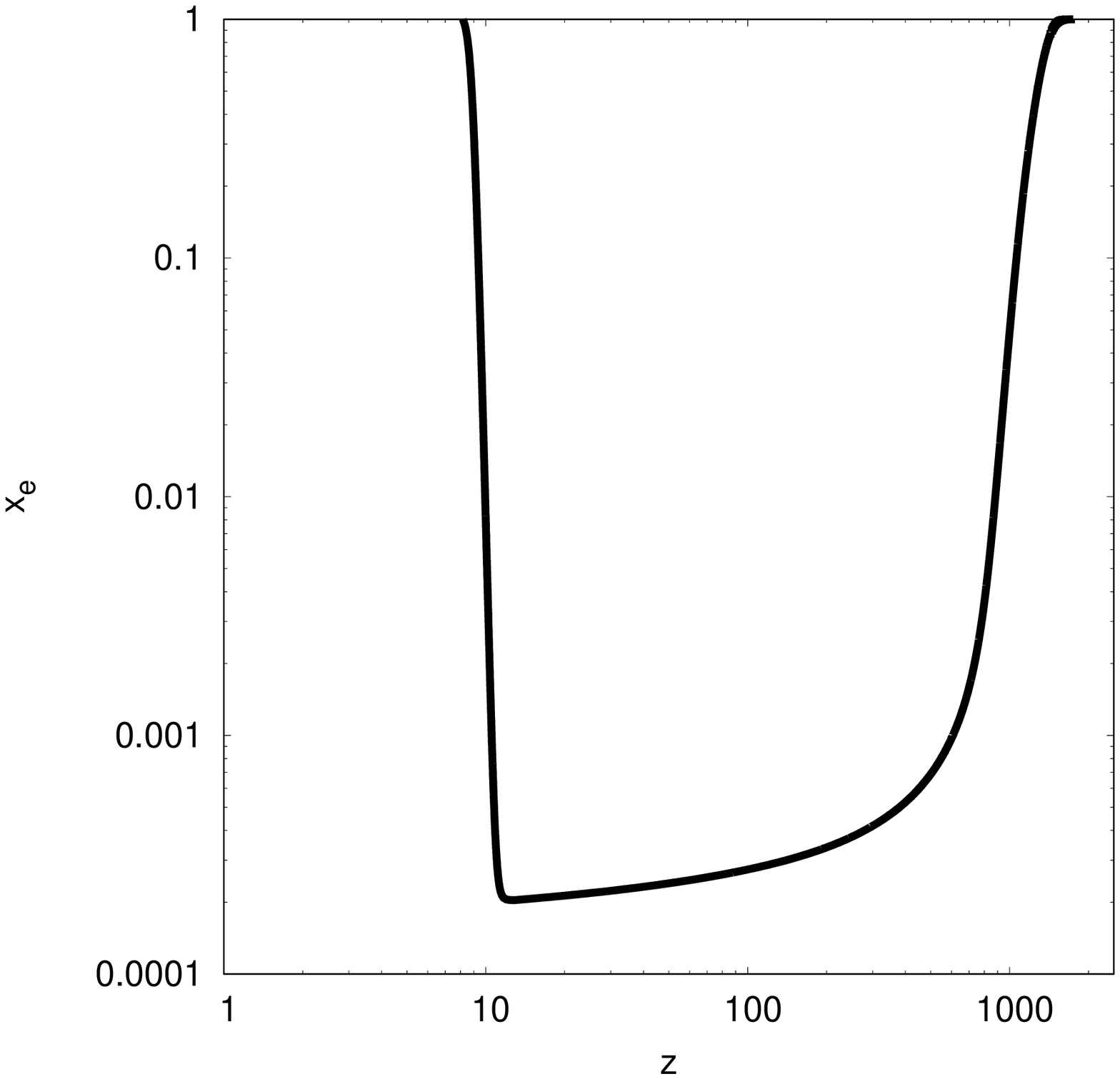}
}
\caption{The optical depth ({\it left}), the visibility function ({\it middle}) and the ionization fraction ({\it right}) for the bestfit parameters from Planck 15 (TT,TE,EE+lowP+lensing+ext(BAO+JLA+$H_0$)).}
\label{fig:visibility}
\end{figure}

\begin{figure}
\centering{\includegraphics[width=0.55\linewidth]{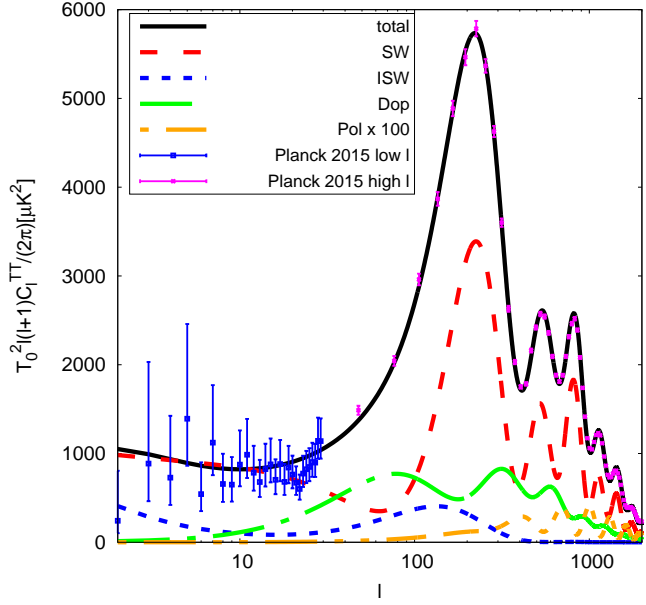}}
\caption{The angular power spectrum of the temperature autocorrelation function calculated for the bestfit parameters from Planck 15 (TT,TE,EE+lowP+lensing+ext(BAO+JLA+$H_0$)). Also shown are the binned Planck 2015 data \cite{planck15-ps}. The different contributions due to the Sachs-Wolfe effect ("SW"), the  integrated Sachs-Wolfe effect ("ISW"), the Doppler term ("Dop") as well as the (amplified) polarization term ("Pol x 100") are also included.}
\label{fig:fig6}
\end{figure}

The Boltzmann hierarchy is solved by a line-of-sight integral \cite{Seljak:1996is, Hu:1997hp} which for the brightness perturbation of the scalar mode is given by
\begin{eqnarray}
\frac{\Theta_{\ell}}{2\ell +1}=\int_{\eta_{in}}^{\eta_0}d\eta S(k,\eta)j_{\ell}\left[k\left(\eta_0-\eta\right)\right],
\end{eqnarray}
where $j_{\ell}(x)$ denote the spherical Bessel functions and $S(k,\eta)$ is the source function which can be written in terms of different physical contributions $S_i$ as (e.g., \cite{Lesgourgues:2013bra, durrer})
\begin{eqnarray}
S&=&\sum_iS_i.
\end{eqnarray}
The sum includes 
the Sachs-Wolfe term which is important at low $\ell$, corresponding to large angular scales,
\begin{eqnarray}
S_{SW}=\frac{g}{4}\Delta_{\gamma},
\end{eqnarray}
the integrated Sachs-Wolfe term,
\begin{eqnarray}
S_{ISW}=g(\Phi+\Psi)-e^{-\tau}(\dot{\Phi}+\dot{\Psi}),
\end{eqnarray}
the Doppler term involving the gauge-invariant baryon velocity $V_b$
\begin{eqnarray}
S_{dop}=k^{-1}\left(g\dot{V}_b+\dot{g}V_b\right)
\end{eqnarray}  
and a polarization term  
\begin{eqnarray}
S_{pol}=gP^{(0)}
\end{eqnarray}
involving the quadrupole of the brightness perturbation as well as of the E-mode polarization. 
Moreover, $g\equiv\dot{\tau}e^{-\tau}$ is the visibility function and $\tau(t)=\sigma_T\int_t^{t_0}n_e(t)dt$
is the optical depth. The term $e^{-\tau(t)}$ gives the probability that a CMB photon observed today, that is, at $t_0$,
 has not scattered since time $t$. The  optical depth and the visibility function are shown in figure \ref{fig:visibility}. 
 All figures in this section have been done using the bestfit parameters of the $\Lambda$CDM base model derived from Planck 15 temperature, polarization data, lensing likelihood and additional data from large scale structure and supernovae (BAO, JLA and $H_0$) at 68 \%CL
 given by $\Omega_bh^2=0.02230\pm 0.00014$, $\Omega_c h^2=0.1188\pm 0.0010$, $z_{re}=8.8^{+1.2}_{-1.1}$, $A_s=(2.142\pm 0.049)\times 10^{-9}$, $n_s=0.9667\pm 0.0040$ \cite{planck15-cosmo}.
As can be seen it is strongly peaked around decoupling at around $z=1089.90$. Upto a redshift around $z=8.8$ the visibility function decays monotonously and then rises again due to the complete reionization of the universe at late times. 
This will  be discussed in more detail below.
The different contributions to the total angular power spectrum of the brightness perturbation induced by the adiabatic, scalar mode are shown in figure \ref{fig:fig6}.
Also shown are the binned Planck 2015 data. As can be appreciated from figure \ref{fig:fig6} different contributions are important on different scales.
On large scales the Sachs-Wolfe plateau can be identified. 
The subsequent  overall peak structure are an imprint of the acoustic oscillations of the baryon-photon fluid in the tight coupling regime, long before last scattering of the photons.
This can be seen from the evolution equations of the density contrast and velocities of baryons and photons in the tight-coupling limit  for the scalar mode, which are given  in the gauge-invariant formalism (cf. \cite{KS}),  
  \begin{eqnarray}
  \dot{\Delta}_{\gamma}&=&-\frac{4}{3}kV_{\gamma},\\
  \dot{V}_{\gamma}&=&k(\Psi-\Phi)+\frac{k}{4}\Delta_{\gamma}-\frac{k}{6}\pi_{\gamma}+\tau_c^{-1}(V_b-V_{\gamma})
  \end{eqnarray}
  and the baryons satisfy,
  \begin{eqnarray}
  \dot{\Delta}_b&=&-kV_b-3c_s^2{\cal H}\Delta_b,\\
  \dot{V}_b&=&(3c_s^2-1){\cal H}V_b+k(\Psi-3c_s^2\Phi)+kc_s^2\Delta_b+R\tau_c^{-1}(V_{\gamma}-V_b),
  \end{eqnarray}
 where $\tau_c^{-1}=an_e\sigma_T$ is the mean free path of photons between scatterings with $\sigma_T$ the Thomson cross section. Moreover, $R\equiv\frac{4}{3}\frac{\rho_{\gamma}}{\rho_b}$ and the $c_s^2=\frac{\partial\overline{p}}{\partial\overline{\rho}}$ is the adiabatic sound speed calculated in terms of the background quantities.
 Here the dot denotes the time derivative w.r.t. to conformal time $\eta$ and ${\cal H}=\frac{\dot{a}}{a}$.
 These equations can be combined to yield a second order differential equation for $\Delta_{\gamma}$ which describes a forced harmonic oscillator \cite{Hu:1994uz},
 \begin{eqnarray}
 \ddot{\Delta}_{\gamma}+\frac{\dot{R}_b}{1+R_b}\dot{\Delta}_{\gamma}+c^2_{sb\gamma}k^2\Delta_{\gamma}
 \simeq\frac{4k^2}{3}\frac{2+R_b}{1+R_b}\Phi,
 \label{eq:acoustic}
 \end{eqnarray}
 where $R_b\equiv\frac{1}{R}$ and $c^2_{sb\gamma}\equiv\frac{1}{3}\frac{1}{R_b+1}$ is the sound speed of the baryon-photon fluid. For adiabatic initial conditions  this is solved by
 \begin{eqnarray}
 \Delta_{\gamma}(\eta)&=&\frac{1}{(1+R_b)^{\frac{1}{4}}}\left[\Delta_{\gamma}(0)\cos(k r_s(\eta))
 +\frac{\sqrt{3}}{k}\left[\dot{\Delta}_{\gamma}(0)+\frac{1}{4}\dot{R}_b(0)\Delta_{\gamma}(0)\right]\sin(kr_s(\eta))
 \right.
 \nonumber\\
 &&\left.
 +\frac{\sqrt{3}}{k}\int_0^{\eta}d\tau'\left(1+R_b(\eta')\right)^{\frac{3}{4}}\sin\left[kr_s(\eta)-kr_s(\eta')\right]F(\eta')
 \right]
 \end{eqnarray}
 where $F(\eta)\equiv\frac{4k^2}{3}\frac{2+R_b}{1+R_b}\Phi$ 
 and the sound horizon is defined by
 \begin{eqnarray}
 r_s(\eta)\equiv\int_0^{\eta}c_{sb\gamma}d\eta'.
 \end{eqnarray}
 After the first acoustic peak around $\ell=220$ a clear damping of the amplitude of the subsequent peaks is observed. This is due to the photon diffusion damping. As the universe cools down the number of free electrons drastically diminishes as they combine with the nuclei. This implies that the mean free path of photons increases strongly and tight coupling between the baryon and photon fluids breaks down. However, as this is not an instantaneous process photons diffuse and later on start free streaming within the baryon fluid thereby moving baryons out of the potential wells of the CDM gravitational field. This results in an attenuation of the baryon density perturbation which in turn 
 results in a damping of the corresponding brightness perturbations which manifests itself in the CMB temperature as well as polarization angular power spectra. The resulting damping is clearly seen in the numerical solutions of the Boltzmann hierarchy. However, by  using a simple model of a random walk it can also be estimated and understood physically. It is rather interesting that the effect of the damping is well approximated by simply multiplying the amplitudes by a factor $\exp(-k^2/k_D^2(\eta))$ 
 as the comparison between analytical approximations and  numerical solutions shows \cite{Hu:1994uz}.
 Here, $k_D^{-1}$, is the Silk scale or photon diffusion scale. It is roughly the comoving distance a photon can travel since some initial time. It can be estimated by modelling the movement of the photon in the local baryon rest frame as a random walk (e.g., \cite{Lyth:2009zz}). In this case the mean time between collisions (i.e. scattering off) with electrons is 
 $t_c\sim (\sigma_T n_e)^{-1}$. The average number of steps during a time interval $t$ is thus $N=t/t_c$.
 Therefore, during a time interval $t$ the photon diffuses a distance of the order $d\sim\sqrt{N}t_c\sim(tt_c)^{\frac{1}{2}}$. This yields the physical photon diffusion scale
 \begin{eqnarray}
 ak_D^{-1}\simeq\left(\frac{t}{\sigma_T n_e}\right)^{\frac{1}{2}}.
 \end{eqnarray}
 Therefore, the comoving photon diffusion scale evolves with the scale factor during radiation domination as
 $k_D^{-1}\sim a^{3/2}$ and during matter domination as
 $k_D^{-1}\sim a^{5/4}$.
  Each scale $k^{-1}$ starts out bigger than the Silk scale at horizon entry. Once $k_D(\eta)=k$ is reached photon diffusion damping sets in.
  
 The amplitude of the angular power spectra depends rather strongly on the cosmological parameters. For example, whereas increasing the baryon energy density parameter strongly increases the first acoustic peak, increasing the total matter energy density lowers the first peak.

 Another important observation of the CMB is its linear polarization.
 This polarization is due to Thomson scattering of photons off free electrons at around last scattering. 
 The polarization of  a plane electromagnetic wave   is commonly determined by the behaviour of  its electric field. 
The magnetic field is obtained from  Maxwell's equations.
 The most general form of a homogeneous monochromatic plane wave travelling in the direction 
 ${\bm k}=k\hat{\bm n}$ is given by
(cf. e.g. \cite{jackson}),
 \begin{eqnarray}
 {\bm E}=\left({\bm\epsilon}_1E_1+{\bm\epsilon}_2E_2\right)e^{i\left({\bm k}\cdot{\bm x}-\omega t\right)}
 \end{eqnarray}
 where ${\bm \epsilon}_i$  are the polarization vectors which together with $\hat{\bm n}$ can be chosen to form a right handed orthonormal basis.  
 The amplitudes $E_i$ can be conveniently written separating out the phase $\delta_i$ so that $E_i=s_ie^{i\delta_i}$, $i=1,2$. The wave is linearly polarized if there is no phase difference, i.e. $\delta_1=\delta_2$. In this case  
 the electric field has a constant direction which makes an  angle $\theta=\tan^{-1}\left(\frac{E_2}{E_1}\right)$ with the direction ${\bm\epsilon}_1$. Otherwise, in general,  the wave is elliptically polarized with the particular case of circular polarization for which the amplitudes satisfy $|E_1|=|E_2|$ and the phase difference is $|\delta_1-\delta_2|=\frac{\pi}{2}$.
 There is another choice of basis vectors instead of those spanning  the plane transverse to the direction of propagation,  which is the helicity basis with the basis vectors ${\bm \epsilon}_{\pm}=\frac{1}{\sqrt{2}}\left({\bm\epsilon}_1\pm i{\bm\epsilon}_2\right)$. 
 The description of radiation requires in general four parameters, namely, its intensity, degree of polarization, plane of polarization and the ellipticity of radiation. These distinct parameters are encoded in a compact way in the four Stokes parameters (cf. e.g., \cite{chandra}).
 The Stokes parameters in terms of the linear polarization basis  vectors ${\bm\epsilon}_{1,2}$ are given by (e.g., \cite{jackson})
 \begin{eqnarray}
 I&=&\left|{\bm\epsilon}_1\cdot{\bm E}\right|^2+\left|{\bm\epsilon}_2\cdot{\bm E}\right|^2\Rightarrow I=s_1^2+s_2^2
 \label{eq:s1}
 \\
 Q&=&\left|{\bm\epsilon}_1\cdot{\bm E}\right|^2-\left|{\bm\epsilon}_2\cdot{\bm E}\right|^2\Rightarrow Q=s_1^2-s_2^2
\label{eq:s2}\\
U&=&2{\rm Re}\left[({\bm \epsilon}_1\cdot{\bm E})^*({\bm \epsilon}_2\cdot{\bm E})\right]\Rightarrow U=2s_1s_2\cos(\delta_2-\delta_1)
\label{eq:s3}\\
V&=&2{\rm Im}\left[({\bm \epsilon}_1\cdot{\bm E})^*({\bm \epsilon}_2\cdot{\bm E})\right]\Rightarrow V=2s_1s_2\sin(\delta_2-\delta_1)
 \label{eq:s4}
 \end{eqnarray}
 These satisfy, 
 \begin{eqnarray}
 I^2=Q^2+U^2+V^2.
 \label{eq:fullypol}
 \end{eqnarray}
  Moreover, going to the helicity basis it can be shown that $V=s_+^2-s_{-}^2$ giving the difference between the intensities of positive and negative helicity states. 
From equations (\ref{eq:s1}) to (\ref{eq:s4}) it follows that for linearly polarized radiation, $Q\neq 0$, $U\neq 0$ but  
 $V=0$. 
For circularly polarized light, Q=0=U and $V\neq0$.
So far we have considered the case of one monochromatic plane wave. However, in general in astrophysics and cosmology one would  rather expect a diffuse, not monochromatic radiation which is not 100\% polarized  but rather partially polarized. This type of radiation can be described by the incoherent sum of unpolarized and fully polarized contributions leading to $Q^2+U^2+V^2<I^2$. This leads to  the definition of the degree of polarization $p$,
\begin{eqnarray}
p=\frac{\sqrt{Q^2+U^2+V^2}}{I}
\end{eqnarray}
This reduces to the definition of the   degree of linear polarization for $V=0$ \cite{Tinbergen}.

  Rotating the  linear polarization vectors ${\bm\epsilon}_{1,2}$ by an angle $\psi$ around the normal $\hat{\bm n}$, namely, (cf. e.g. \cite{PU})
 \begin{eqnarray}
 {\bm\epsilon}_1'&=&\cos\psi{\bm\epsilon}_1+\sin\psi{\bm\epsilon}_2, 
 \nonumber\\
 {\bm\epsilon}_2'&=&-\sin\psi{\bm\epsilon}_1+\cos\psi{\bm\epsilon}_2
 \end{eqnarray}
 leaves $I$ and $V$ invariant while the combination $Q\pm iU$ behaves as a spin-2 variable, i.e.
 \begin{eqnarray}
 (Q\pm iU)'=e^{\mp 2i\psi}(Q\pm iU).
 \label{eq:rot}
 \end{eqnarray}
 This makes this combination to be the most useful one  to describe the polarization of the CMB.
 Whereas the temperature fluctuations or brightness perturbations are expanded in spin-0 spherical  harmonics $Y_{\ell m}$  (cf. equation ({\ref{theta})) the CMB polarization variables are expanded in terms of spin-2 spherical harmonics
 $_{\pm 2}Y_{\ell m}$ such that \cite{HW}
 \begin{eqnarray}
 (Q\pm iU)(\eta,{\bm x},\hat{\bm n})=\int\frac{d^3k}{(2\pi)^3}\sum_{\ell}\sum_{m=-2}^{2}
 (E_{\ell}^{(m)}\pm iB_{\ell}^{(m)})\, _{\pm 2}G^m_{\ell},
 \end{eqnarray}
 where $_{\pm 2}G^m_{\ell}=(-1)^{\ell}\sqrt{\frac{4\pi}{2\ell+1}}\, _{\pm 2}Y^{m}_{\ell}(\hat{\bm n})
 e^{i{\bm k}\cdot{\bm x}}$. As before, scalar modes correspond to $m=0$, vector modes  to $m=\pm 1$ and tensor modes to $m=\pm 2$.
 $E_{\ell}^{(m)}$ and $B_{\ell}^{(m)}$ are the E-mode and B-mode of polarization. By using a representation of the spin-2 spherical harmonics in terms of derivative operators it can be shown that these modes can be written in terms of a gradient and a curl of a field, respectively. This gave rise to the notation in terms of E-mode ({\it ``electric''}) and B-mode ({\it ``magnetic''}).
 
Before photon decoupling the tight coupling of the baryon-photon fluid is achieved by Thomson scattering.
Its cross section is given by
\begin{eqnarray}
\frac{d\sigma}{d\Omega}\left({\bm n},{\bm\epsilon} ; {\bm n}^{(0)}, {\epsilon}^{(0)}\right)
=\frac{3\sigma_T}{8\pi}\left|{\bm\epsilon^{*}}\cdot{\bm\epsilon^{(0)}}\right|^2
\label{ThomsonXsect}
\end{eqnarray}
where ${\bm \epsilon^{(0)}}$ and ${\bm\epsilon}$ are the polarization vectors of the incoming and outgoing photons, respectively. The Thomson cross section is given by $\sigma_T\equiv\frac{8\pi}{3}\left(\frac{e^2}{m_ec^2}\right)^2$.
The incoming and outgoing propagation directions of the photons are ${\bm n}^{(0)}$ and ${\bm n}$, respectively.
Thomson scattering does not source circular polarization. Therefore in the $\Lambda$CDM model $V=0$ and generally it is not considered in CMB physics though there are some exceptions.
The two polarization vectors of the incoming photons can be chosen in spherical coordinates as (cf. e.g. \cite{jackson})
 \begin{eqnarray}
 {\bm\epsilon}_1^{(0)}&=&\cos\theta\left({\bm e}_x\cos\phi+{\bm e}_y\sin\phi\right)-{\bm e}_z\sin\theta
 \label{pol1}\\
 {\bm\epsilon}_2^{(0)}&=&-{\bm e}_x\sin\phi+{\bm e}_y\cos\phi.
 \label{pol2}
 \end{eqnarray}
 The polarization vectors of the outgoing (scattered) radiation will be chosen along the cartesian coordinate ${\bm x}$- and ${\bm y}$-axes, respectively, so that
 \begin{eqnarray}
 {\bm\epsilon}_1&=&{\bm e}_x
 \label{pol3}\\
 {\bm\epsilon}_2&=&{\bm e}_y
 \label{pol4}
 \end{eqnarray}
 Moreover ${\bm n}^{(0)}$ and ${\bm n}$ make an angle $\theta$.
 The differential cross section by definition is the ratio of the  power radiated in the direction ${\bm n}$ with polarization ${\bm\epsilon}$ per unit solid angle and the unit incident flux which is the power per unit area in the direction ${\bm n}_0$ and polarization ${\bm\epsilon}_0$. Thus  for scattering of electromagnetic waves off one electron at a distance $r$ from the electron \cite{jackson}, 
 \begin{eqnarray}
 \frac{d\sigma}{d\Omega}\left({\bm n},{\bm\epsilon} ; {\bm n}^{(0)}, {\epsilon}^{(0)}\right)=
 \frac{r^2\frac{1}{2}\left|{\bm\epsilon^*}\cdot{\bm E}\right|^2}
 {\frac{1}{2}\left|{\bm\epsilon}^{(0)*}\cdot{\bm E}^{(0)}\right|^2}
\label{diffXsect}
 \end{eqnarray}
 where ${\bm E}^{(0)}$ and ${\bm E}$ are the electric fields of the incoming and scattered radiation far away from the scatterer, respectively.
 Therefore using equations (\ref{ThomsonXsect}) and (\ref{diffXsect}) and defining
 partial intensities $I_m\equiv\left|{\bm\epsilon}^*_m\cdot{\bm E}\right|^2$ and equivalently for the incident radiation, then
 \begin{eqnarray}
 I_m=r^{-2}\sum_{n=1,2}\left(\frac{d\sigma}{d\Omega}\right)\left(\hat{\bm n},{\bm\epsilon}_m; \hat{\bm n}^{(0)},{\bm\epsilon}^{(0)}_n\right)I^{(0)}_n.
 \end{eqnarray}
 Assuming that the incident radiation is unpolarized inplies that the intensity only depends on the propagation direction 
${\bm n}^{(0)}$ and not on the polarization vectors thus $I_n^{(0)}=I_n^{(0)}(\hat{\bm n}^{(0)})$ (e.g. \cite{dodelson}). Moreover $I_1^{(0)}=\frac{I^{(0)}}{2}=I_2^{(0)}$. Using the expressions (\ref{pol1})-(\ref{pol4}) for the polarization vectors and integrating over all incoming directions ${\bm n}^{(0)}$ the Stokes parameter $Q$ (cf. equation (\ref{eq:s2}))
reads
\begin{eqnarray}
Q=-\frac{3\sigma_T}{16\pi}\int d\Omega_{\bm n^{(0)}}r^{-2}I^{(0)}({\bm n}^{(0)})\cos(2\phi)\sin^2\theta.
\label{eq:Q}
\end{eqnarray}
 The brightness perturbation $\Theta$ is determined by the fractional perturbation of the intensity $I$.
 Using the expansion in terms of spin-0 spherical spherical harmonics $Y^m_{\ell}({\bm n})$ (cf. equation (\ref{theta})) it follows that the polarization parameter $Q$ is determined by the quadrupole moment of the brightness perturbation since 
 $\sin^2\theta\cos 2\phi\propto \left[Y^2_2(\hat{\bm n}^{(0)})+Y^{-2}_2(\hat{\bm n}^{(0)})\right]$. 
 Moreover, since $U(\hat{\bm n}^{(0)})$ can be obtained from $Q(\hat{\bm n}^{(0)})$  by a coordinate transformation $\phi\rightarrow\phi-\frac{\pi}{4}$  (cf. equation (\ref{eq:rot}))
 the corresponding expression for $U$ is simply given by equation (\ref{eq:Q}) with $\cos(2\phi)$ replaced by $\sin(2\phi)$. Therefore, $U$ is also sourced by the quadrupole of the brightness perturbation since $\sin^2\theta\propto\left[ Y^2_2(\hat{\bm n}^{(0)})-Y_2^{-2}(\hat{\bm n}^{(0)})\right]$.

 The evolution of the brightness perturbations  indicate that only close to photon decoupling the amplitude of the multipoles can grow significantly. Thus the primordial polarization is generated at last scattering.
 Since it is sourced by the brightness perturbation it also shows the characteristic oscillating structure 
 imprinted by the oscillations of the baryon photon fluid (cf figure \ref{fig:fig7}). 
  \begin{figure}
\centering{\includegraphics[width=0.45\linewidth]{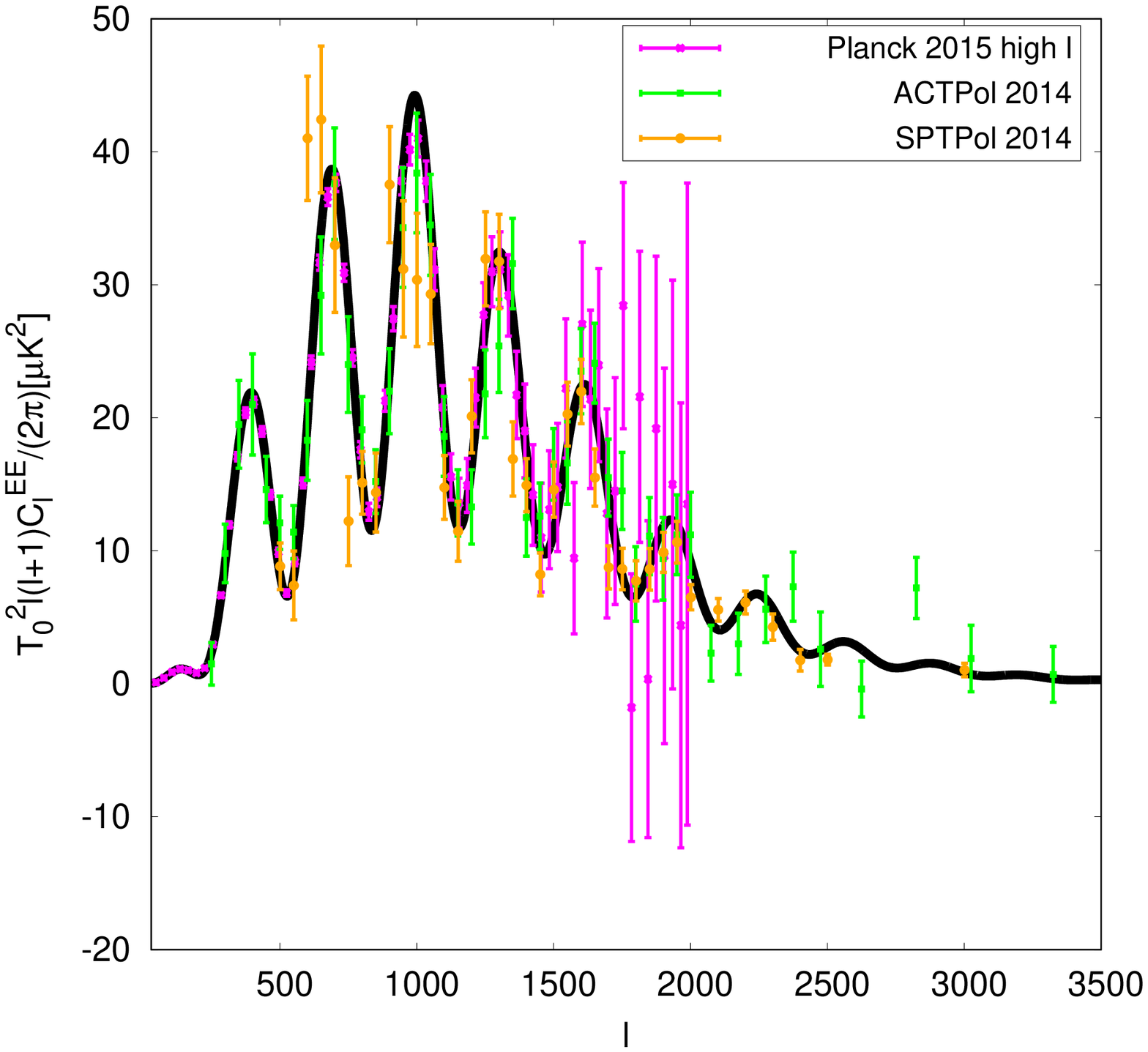}\hspace{1.1cm}
\includegraphics[width=0.45\linewidth]{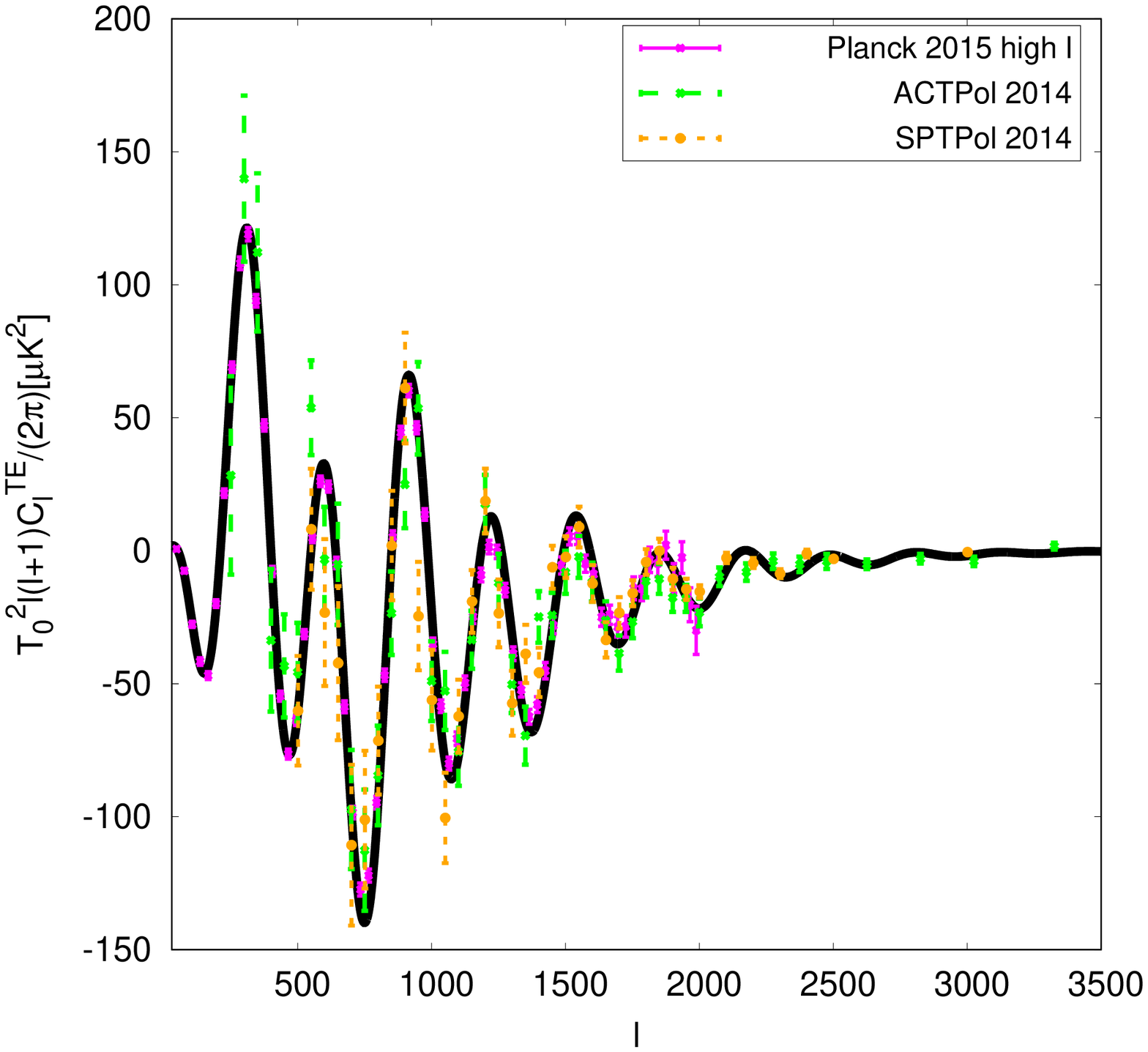}}
\caption{The angular power spectrum of the E-mode autocorrelation function  ({\it left}) and the cross correlation function of the temperature and E-mode ({\it right}) for the bestfit parameters from Planck 15 (TT,TE,EE+lowP+lensing+ext(BAO+JLA+$H_0$)) together with the Planck 2015 \cite{planck15-ps}, ACTPol \cite{actpol-Naess} and SPTPol \cite{sptpol-Crites} data.}
\label{fig:fig7}
\end{figure}
 Six acoustic peaks in the E-mode autocorrelation functions have been observed by ACTPol \cite{actpol-Naess} and SPTPol \cite{sptpol-Crites}. 
 The feature at low ${\ell}$ in the angular power spectrum of the auto- and temperature cross correlation function of the $E$- mode (cf. figure \ref{fig:fig7}) is due to reionization at low redshifts. In the reionized universe the number of free electrons is very large again. Some of the CMB radiation gets re-scattered thereby increasing the polarization signal.

 The first observations that the universe today is completely ionized came from observations of quasar spectra.
 The intergalactic medium (IGM) which permeates the space between galaxies  can be probed by its emission and absorption spectra of light from distant background sources. 
 Quasars are ideal background sources since due to their immense power they can be observed upto very high redshifts. Hydrogen is the most abundant element in the universe. 
 If hydrogen is neutral in regions along the line of sight then the 
 Lyman $\alpha$ absorption at the corresponding rest frame wavelengths should lead to the 
 "Gunn-Peterson trough"  in the quasar spectrum at wavelengths shorter than the terrestrial reference wavelength at $\lambda_{{\rm Ly}\alpha}=121.6$ nm. For objects at redshifts beyond the reionization redshift 
 a complete Gunn-Peterson trough should be observed. Discrete absorption features are observed due to residual regions of neutral hydrogen which give rise to the so-called Lyman $\alpha$ forrest seen in quasar spectra.
With  surveys such as the Sloan Digital Sky Survey (SDSS) a number of quasars at redshifts $z\stackrel{\sim}{>} 6$ have been observed. Their spectra show indications for a complete Gunn-Peterson trough and thus imply a lower limit on the reizonation redshift at around $z\simeq 6$.
 Observations of the CMB are well fitted by assuming instanteneous reionization at a redshift $z_{reio}$.
A typical evolution of the ionization fraction is shown in figure \ref{fig:visibility}.
Planck 15 data are bestfit for a reionization redshift $z_{reio}=8.8^{+1.2}_{-1.1}$ at 68\%CL (using the temperature and low $\ell$ polarization data and lensing reconstruction including external data from BAO and supernovae observations) \cite{planck15-cosmo}.
The origin of this reionization of the IGM lies with the formation of first objects. The first generation of stars, the population III stars, are assumed to be very hot and massive. Their radiation ionizes the intergalactic gas in their vicinity and thus play an important role in the reionization of the universe.

The absolute spectrum of the CMB offers another possibility to gain insight into the physical conditions of the universe during its evolution. The COBE/FIRAS instrument measured the absolute spectrum of the CMB \cite{COBE}.
The data analysis showed that the CMB is very close to a perfect black body spectrum.
The deviations from a Planck spectrum are very small indeed. 
These spectral distortions are classified in  two types, namely, $\mu$-type and $y$-type. They are generated at different times during the evolution of the universe as 
 the spectral properties depend on the interactions of photons with the cosmic plasma.
 Deviations from the Planck spectrum result when energy is injected into the photon distribution or the photon occupation number density is changed. This could be caused, e.g., by particle-antiparticle annihilation reactions
 $X\overline{X}\rightarrow\gamma\gamma$ or particle decay $X\rightarrow ...\gamma$.
 Other interesting processes include the damping of density perturbations (cf. the Silk damping discussed above) \cite{Hu:1994bz}
 or damping of primordial magnetic fields \cite{jko,koku}.
 If this happens at very early times, at very high energies, corresponding to redshifts $z>2\times 10^6$ then bremsstrahlung and double Compton scattering are very efficient \cite{Hu:1992dc}. Since these interactions do not conserve photon numbers and redistribute photon frequencies a Planck spectrum can always be re-established. At later times and thus lower temperatures these interactions are no longer efficient. If the photon energy and/or occupation numbers are suddenly changed for redshifts between $2\times 10^6$ and $4\times 10^4$ photons interact with electrons
 by elastic Compton scattering. This changes the photon frequency distribution but not the occupation numbers
 resulting in a deviation from the Planck spectrum in the form of a Bose-Einstein spectrum,
 $f=\left[\exp\left(\mu+h\nu/kT_e\right)-1\right]^{-1}$. This gives rise to a $\mu$-type spectral distortion.
 For redshifts less than $5\times 10^4$ elastic Compton scattering becomes inefficient. Any energy injection at this later times result in a $y$-type spectral distortion encoded by expanding the spectrum around the Planck spectrum
$B_{\nu}(T)$ as 
 \begin{eqnarray}
 B_{\nu}(T)+y\frac{\partial S_y}{\partial y},
 \hspace{2cm}
 y=\int\frac{k\left(T_e-T_{\gamma}\right)}{m_ec^2}d\tau
 \end{eqnarray}
where $T_e$ is the electron (matter) temperature and $T_{\gamma}$ the photon temperature. $\tau$ is the optical depth (as defined above).
This separation  into epochs when either $\mu$-type or $y-$type spectral distortions are generated has been re-fined by including an intermediate era generating an $i-$type spectral distortion \cite{Khatri:2012tw}.
The observational limits of COBE/FIRAS are $|\mu|<9\times 10^{-5}$ and $|y|<1.5\times 10^{-5}$ at 95\% CL.
There are discussions about future missions, such as the Primordial Inflation Explorer PIXIE \cite{pixie}, a potential NASA mission, to improve the COBE/FIRAS measurements of the absolute spectrum of the CMB.

 The matter power spectrum provides the initial conditions for large scale structure formation.
 Linear density perturbations at some point reach a critical amplitude and enter 
 the non-linear regime resulting in gravitational collapse and giving rise on larger scales to the nontrivial network of voids, sheets and filaments as observed in galaxy surveys as well as numerical simulations.
 The total matter power perturbation $\Delta_m$ has a contribution from the perturbation of the cold dark matter component $\Delta_c$  as well as in the baryons $\Delta_b$,
 \begin{eqnarray}
 \Delta_m\equiv R_c\Delta_c+R_b\Delta_b,
 \end{eqnarray}
 where during the matter dominated era $R_i\equiv\frac{\rho_i}{\rho_{matter}}$, $i=b,c$.
 The corresponding matter spectrum is given by 
 \begin{eqnarray}
 P_m(k)=\frac{k^3}{2\pi^2}|\Delta_m|^2.
 \end{eqnarray}
  The total linear matter spectrum for the adiabatic mode  is shown in figure \ref{fig-matterps}.
  \begin{figure}
\centering{\includegraphics[width=0.45\linewidth]{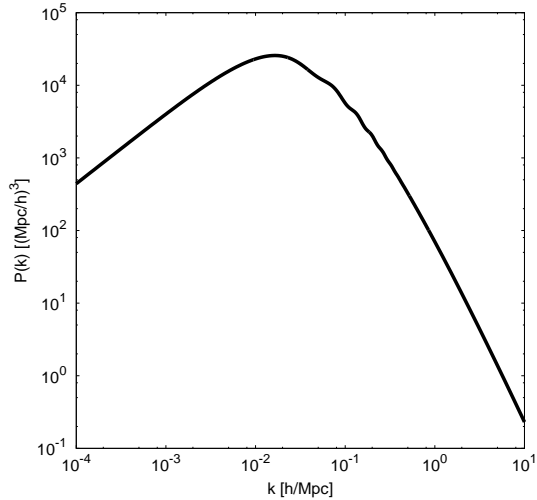}}
\caption{The total linear matter power spectrum for the bestfit parameters from Planck 15 (TT,TE,EE+lowP+lensing+ext(BAO+JLA+$H_0$)).}
\label{fig-matterps}
\end{figure}
  The evolution of the matter perturbation is determined in general 
  by the continuity equation, (cf., e.g., \cite{CoorayShethPhysRep})
  \begin{eqnarray}
  \frac{\partial\Delta}{\partial t}+\frac{1}{a}{\bf\nabla}\cdot(1+\Delta){\bf u}=0
  \end{eqnarray}
  and the Euler equation 
  \begin{eqnarray}
  \frac{\partial\Delta}{\partial t}+H{\bf u}+\frac{1}{a}\left[\left({\bf u}\cdot{\bf\nabla}\right){\bf u}
  +{\bf\nabla}\phi\right]=0
  \end{eqnarray}
  where ${\bf u}={\bf v}-H{\bf x}$ is the peculiar velocity.
Moreover,  the gravitational potential $\phi$ is determined by the Poisson equation
  \begin{eqnarray}
  \nabla^2\phi=4\pi G\rho a^2\Delta.
  \end{eqnarray}
  In the linear regime the equations of linear cosmological perturbation theory are recovered on sub horizon scales. Beyond the linear regime there are approaches to treat the evolution in a semianalytical way such as the Zeldovich approximation or the adhesion model. However, in order to obtain a complete picture of the evolution of the gravitational instabilities and subsequently the distribution of mass in the universe it is necessary to ressort to numerical simulations. One of the largest simulations done is the Millenium Run of the Virgo Consortium \cite{Springel:2005nw} which uses the publicly available {\tt GADGET-2} code \cite{gadget-2}.  These gravitational N-body simulations generally confirm the picture of how the complex web of cosmic voids and filaments forms from the initial,  linear density perturbations.

 \section{Dark matter and dark energy}
 \label{sec4}

 Determining the velocity dispersions of galaxies in clusters Fritz  Zwicky  found the first evidence  already in 1933 that there is more matter in the universe than is visible. Typically in spiral galaxies and galaxy clusters the member objects such as stars, gas clouds or galaxies follow a circular movement around the center of the corresponding structure. 
Plotting this radial dependence of the velocities yield the rotation curves.
These show a flattening at large distances from the center which is an indication of  the presence of non luminous matter.
The rotational velocities can be estimated from the Doppler shift of spectral line(s) in the corresponding spectra. 
In particular, in galaxies the presence of neutral hydrogen gas clouds (HI clouds) allows to use the Doppler shifts of their 21cm line emission to determine the rotation velocities. 
 The 21cm line or 1420 MHz radiation emission  is due to the transition between the two hyperfine levels of the 1s ground state of hydrogen. This transition is highly forbidden with a transition probability of $2.6\times 10^{-15}$ s$^{-1}$. However, there is a significant signal due to the large amount of hydrogen in the universe. Observations of the 21 cm line emission have been an important tool in galactic astronomy. It is becoming now the future of cosmological observations to explore the universe 
 at high redshifts before reionization during the so called dark ages. These refer to the epoch after last scattering and before  the formation of the first luminous objects such as the first generation stars (cf. e.g. \cite{21cm-cosmo}).

The dynamics can be described by a rather simple model which allows to get an estimate of the total mass of the structure (cf., e.g.,  the description in 
\cite{Perkins:2003pp, Coles:1995bd}).
 The rotation curves determine the rotational velocity $v_{rot}(r)$ as a function of radial distance $r$ from the center. Thus applying Newtonian dynamics yields to,
\begin{eqnarray}
\frac{GM_r}{r^2}=\frac{v_{rot}^2(r)}{r}\rightarrow
M_r=\frac{v^2_{rot}(r)r}{G},
\end{eqnarray}
where $M_r$ is the mass within a radius $r$.
The luminous part of matter in spiral galaxies traces out a central region which has spherical geometry and a disk where the spiral arms are located.  For a star in the central region one expects a linear evolution with radial distance of the rotational
 velocity. This uses the observation that typically in the central region the matter density is constant, thus 
 $M_r\propto r^3$ and
 \begin{eqnarray}
 \frac{Gr^3}{r^2}=\frac{v_{rot}^2}{r}\Rightarrow
 v_{rot}\sim r.
 \end{eqnarray}
Since the mass of the central region is dominant over the contribution from the disk to a good approximation it can be  assumed that on scales much beyond the central region the mass within a radius $r$ is constant.
Thus for a star or gas cloud outside the central region the rotational velocity behaves as
$v_{rot}\sim r^{-\frac{1}{2}}$. Therefore the velocity should increase close to the center and decrease at large distances from the galactic center.
However, the observed rotation curves show a distinct flattening reaching a plateau at large values of $r$.
This behaviour can be explained  by the presence of a galactic halo. Since it is not observed by electromagnetic radiation but only by its gravitational interaction it has to be made of dark matter.
Observations indicate that the halo mass constitutes 80-90\% of the total mass of a galaxy.
This means that light traces only a small part of matter in galaxies and clusters of galaxies.
Extrapolating to even larger scales, only a small of the total energy density is visible. 
The necessity of dark matter was already encountered in section \ref{sec3} discussing  
the observations of the CMB. The bestfit models clearly require cold dark matter, e.g. the Planck 15 temperature and polarization data  constrain  $\Omega_{CDM}h^2=0.02225\pm 0.00016$ at 68\% CL\cite{planck15-cosmo}. Only including cold dark matter  allows to generate  
the corresponding density perturbations in the photon radiation field to fit the observed CMB temperature 
and polarization anisotropies.

The big question now, of course, is what actually is dark matter. Primordial nucleosynthesis or BBN (big bang nucleosynthesis) predicts that the baryonic matter density parameter is $\Omega_b\simeq 0.05$.
However the luminous part of matter in galaxies and galaxy clusters is estimated to have a density parameter
of the order $\Omega_{lum}\simeq 0.01$.
Therefore it is clear that part of dark matter is baryonic.
Bayonic dark matter could be made of Jupiter-like objects or brown dwarfs. The latter are stars with masses less than 10\% of the solar mass implying that their core temperatures are not high enough to start the nuclear reaction chain which fuels the radiation emitted by  stars. Other possible contributions to dark matter could come from black holes. All these objects are called collectively MACHOs which stands for 
massive astrophysical halo objects. There is observational evidence for MACHOs since they act as gravitational lenses leading to the phenomenon of micro lensing. When a MACHO crosses the line-of-sight between us and a distant star it causes the star to appear brighter.
This is due to the gravitational lens effect of the MACHO, namely,   redirecting  light into the direction of the observer and hence leading to a maximum in brightness.
Observing this characteristic local maximum in the time depend light curves of distant stars could be evidence of a microlensing event.
The MACHO project has identified several such  microlensing events observing stars in the Magellanic Clouds and the Galactic Bulge \cite{macho}.

The dominant part of dark matter in the universe is non baryonic. The standard theory is that these are particles created in the very early universe which are stable enough to survive at least until the present day.
Candidates are massive neutrinos, axions and in general weakly interacting massive particles (WIMPs).
These could be for example supersymmetric partners of the standard model of particle physics.

It is interesting to discuss the case of neutrinos in more detail. There are three families of light neutrinos in the standard model of particle physics. Oscillations between different neutrino flavors indicate limits on mass differences, e.g.,   measuring the flux of $\bar{\nu}_e$ from distant nuclear reactors the bestfit to the KamLAND data is $\Delta m^2=6.9\times 10^{-5}$eV$^2$  for a mixing angle $\theta$ determined by $\sin^2 2\theta=0.91$ \cite{kamland}. A different bound on the masses of neutrinos can be obtained from cosmology as follows. Within standard big bang  cosmology light neutrinos
decouple at around 1 MeV as was described above. At this time  neutrinos are relativistic implying that their temperature after decoupling from the rest of the cosmic plasma evolves as $T\propto 1/a$. At the time of neutrino decoupling, photons, neutrinos and the rest of matter which was in thermal equilibrium have the same  temperature. Upto temperatures $T_{m_e}\simeq m_e\sim 0.5$ MeV the temperatures of photons and neutrinos are the same. However, below these temperatures electron positron pair annihilate thereby reducing the number of relativistic degrees of freedom $g_*$. Before electron-positron annihilation the number of relativistic degrees of freedom is determined by photons, electrons and positrons, namely (cf. \cite{KT, Perkins:2003pp})
\begin{eqnarray}
g_{*}(\gamma, e^+, e^{-})=2+\frac{7}{8}\times 4=\frac{11}{2}.
\end{eqnarray} 
At temperatures much below $T_{m_e}$ only the photons contribute to $g_*$, i.e. $g_*(\gamma)=2.$
The entropy in a comoving volume is conserved implying that
\begin{eqnarray}
S=g_*(Ta)^3=const.
\end{eqnarray} 
This yields to 
\begin{eqnarray}
\frac{(Ta)^3_{T<T_{m_e}}}{(Ta)^3_{T>T_{m_e}}}=\frac{g_{*,T>T_{m_e}}}{g_{*,T<T_{m_e}}}=\frac{11}{4}.
\label{eq:gstar}
\end{eqnarray}
As the temperature of the neutrinos evolves as $1/a$ it follows that $(aT_{\nu})_{T>T_{m_e}}=
(aT_{\nu})_{T<T_{m_e}}$. Since at temperatures $T>T_{m_e}$, neutrinos and photons have the same temperature, $T_{\nu}=T_{\gamma}$, equation (\ref{eq:gstar}) implies that for $T<T_{m_e}$
\begin{eqnarray}
\frac{T_{\gamma}}{T_{\nu}}=\left(\frac{11}{4}\right)^{\frac{1}{3}}\simeq 1.4.
\end{eqnarray}
Therefore at present there is a background of (light) neutrinos at a temperature
\begin{eqnarray}
T_{\nu,0}=\left(\frac{11}{4}\right)^{-\frac{1}{3}}T_{\gamma,0}\Rightarrow T_{\nu,0}\simeq 1.95 \;{\rm K}.
\end{eqnarray}
However, so far there has been no direct detection due to the obvious experimental difficulties.
The number density of neutrinos and antineutrinos today is $n_{\nu}=\frac{3}{11}n_{\gamma}$ yielding
$n_{\nu}=113$ cm$^{-3}$. This leads to a bound on the total mass assuming that the neutrino contribution does not overclose the universe, namely satisfying the condition,
$\Omega=\frac{\rho}{\rho_c}\leq 1$. This implies the upper bound
\begin{eqnarray}
\sum_{e,\mu,\tau}m_{\nu}c^2\le 47 \;{\rm eV}.
\end{eqnarray}
Light neutrinos as dark matter constitute hot dark matter since they are relativistic at decoupling ($T\sim 1$ MeV).  However, as they stream freely under the influence of gravity they tend to damp out density perturbations. Therefore the fraction of hot dark matter has to be subdominant w.r.t. cold dark matter.

The number of neutrino species $N_{eff}$ can be constrained with   BBN and CMB data.
For SM neutrinos the relativistic energy density is given by
\begin{eqnarray}
\rho_r=\rho_{\gamma}\left[1+\frac{7}{8}\left(\frac{4}{11}\right)^{\frac{4}{3}}\right].
\end{eqnarray}
A different value for $N_{eff}$ implies a change in the expansion rate and hence the temperature at freeze-out of the conversion reaction between protons and neutrons, $n\leftrightarrow p$. This affects the ratio
$\frac{n}{p}=e^{-\frac{Q}{T_D}}$ and the primordial helium fraction (cf equation (\ref{eq:Yp})).
Recent measurements of the primordial deuterium as well the helium fraction yield \cite{PDG}
\begin{eqnarray}
N_{eff}^{BBN}=2.88\pm 0.16.
\end{eqnarray}

Moreover, changes in  the expansion rate at the time of photon decoupling $t_{dec}$ have implications for  the sound horizon $r_s$ which is  determined by
\begin{eqnarray}
r_s=\int_0^{t_{dec}}c_s\frac{dt}{a}=\int_0^{a_{dec}}\frac{c_s}{a^2H}da
\label{eq:soundh}
\end{eqnarray}
as well as the photon diffusion damping scale $k_D$
\begin{eqnarray}
k_D^{-2}(z)=\int_z^{\infty}\frac{dz}{6H(z)(1+R)\dot{\tau}}\left(\frac{16}{15}+\frac{R^2}{1+R}\right)
\end{eqnarray}
with the differential optical depth $\dot{\tau}=n_e\sigma_T\frac{a}{a_0}$ and the evolution of the baryon over photon fraction $R=\frac{3}{4}\frac{\Omega_{b,0}}{\Omega_{\gamma,0}}(1+z)^{-1}$.
Thus increasing the effective number of relativistic degrees of freedom decreases the damping scale leading to an increase of small scale anisotropies.
Planck 2015 temperature and polarization data  only  \cite{planck15-cosmo} constrain the contribution from light neutrinos and any other dark radiation such that at 68\% CL
$N_{eff}=2.99\pm 0.20$. When combined with large scale structure data this constraint becomes
$N_{eff}=3.04\pm 0.18$.

The experimental evidence of neutrino oscillations implies that neutrinos are massive. The standard $\Lambda$CDM model assumes a mass hierarchy which is dominated by the heaviest neutrino mass eigenstate implying effectively one massive and two massless neutrinos with $\sum m_{\nu}=0.06$ eV. However, a degenerate mass hierarchy is not excluded. The Planck 2015 temperature and polarization data only  \cite{planck15-cosmo} contrain the  sum of the neutrino masses to be $\sum m_{\nu}<0.49$ eV and including large scale structure data leads to the upper bound $\sum m_{\nu}<0.17$ eV at 95\% CL. The absolute neutrino masses can be detected for example using the induced $\beta$ decay resulting in limits on $m(\nu_e)$. Tritium $\beta$ decay experiments such as  KATRIN (e.g., \cite{Drexlin:2013lha}) and PTOLEMY \cite{Betts:2013uya}  are expected to provide the strongest bounds.

The CMB clearly indicates the presence of a substantial amount of cold dark matter. 
The term weakly interacting massive particles (WIMPs) has been introduced to describe 
hypothetical particles which could constitute the cold dark matter in the universe.
These are particles  which are assumed to be non relativistic at the time of their decoupling from the rest of the cosmic plasma. WIMP candidates are, e.g., supersymmetric particles.
The flat FRW model imposes the constraint on the density parameter of any WIMP candidate $\chi$
that   $\Omega_{\chi}<1$ (e.g. \cite{Perkins:2003pp}). 
Since at present WIMPs are non relativistic, the density parameter is determined by 
\begin{eqnarray}
\Omega_{\chi}=\frac{N(T_0)m}{\rho_c}
\end{eqnarray}
assuming a WIMP candidate $\chi$ with mass $m$. Moreover, $N(T_0)=N(T_D)\left(\frac{T_0}{T_D}\right)^3$ is the number density today and $T_0$ and $T_D$ are the temperature today and at "freeze-out" of $\chi\bar{\chi}$ pair production, respectively.
Assuming that freeze-out takes place during the radiation-dominated era the Hubble parameter is given by Eq. (\ref{eq:HRD}) and the expression for the number density of non relativistic especies, Eq. (\ref{eq:nnr}). Moreover at freeze-out the temperature $T_D$ can be estimated by the condition $N(T_D)\langle\sigma v\rangle=H(T_D)$ where $\sigma$ is the $\chi\bar{\chi}$ annihilation cross section and $v$ their relative velocity. Therefore the condition $\Omega_{\chi}<1$ can be used to constrain model parameters.

There are direct as well as indirect experimental searches for WIMP detection within our galaxy. The former one  uses hypothetical WIMP-nucleon scattering within a detector. These signals might have angular and time dependence due to the motion of the earth.
There is a whole range of direct experimental searches going on presently, such as XENON1T\cite{XENON1T} or LUX \cite{LUX}.
Indirect searches use decay products and/or excess radiation of potential WIMP decay or annihilation interactions.
There have been indications of excess $\gamma$ ray radiation from the galactic center from the FERMI satellite (e.g., \cite{FERMI}) as well as positron excess from data of the the AMS \cite{AMS} as well as the HEAT \cite{HEAT} and PAMELA \cite{PAMELA} experiments.

As was already discussed in section \ref{sec2} observations show that at present the largest contribution to the energy density in the universe is due to dark energy. A term which is has been coined  to deal with its unknown origins.
Apart form the observations of supernovae and the CMB anisotropies there is also evidence from large scale structure.
Here we are going to focus on the evidence from baryon acoustic oscillations (BAO) which have been observed in galaxy correlation functions. In section \ref{sec3} the acoustic oscillations in the tightly coupled baryon-photon fluid have been discussed. These give rise to the observed acoustic peak structure in the angular power spectrum of the temperature and polarization anisotropies of the CMB radiation. Moreover, these also manifest themselves as wiggles in the matter power spectrum (cf., Fig. \ref{fig-matterps}). This has been observed, e.g., in the Sloan Digital Sky Survey (SDSS) \cite{Tegmark:2006az, Percival:2007yw}. In real space,   in the two-point galaxy correlation function  these acoustic oscillations in the photon-fluid fluid lead to the already well-detected BAO peak determined by the  size of the sound horizon at decoupling (cf. Eq. (\ref{eq:soundh})), e.g. in the SDSS \cite{Eisenstein:2005su} the BAO peak is detected at $100h^{-1}$ Mpc. This is the largest scale upto which a sound wave can travel before the  sound speed in the baryon fluid decreases dramatically due to the decoupling of the photons.
There are several factors which complicate the detection of the BAO peak  in galaxy redshift surveys due to their low redshift such as the nonlinearity of the density matter field, bias and redshift distortions due to small scale velocities.

The preferred angular separation scale of galaxies marked by the BAO peak can be used as a standard ruler since it is related to the known scale of of the sound horizon at decoupling, $r_s$. This provides a measurement across the line of sight determining $r_s/D_A(z)$ and hence the angular diameter distance $D_A$ as a function of $z$. There is also the possibility to measure the redshift separation along the  line of sight given by $r_s H(z)$. The uncertainty to convert redshifts into distances along and across the line of sight has led to the definition of the parameter \cite{Eisenstein:2005su}
\begin{eqnarray}
D_V(z)=\left[(1+z)^2D_A(z)\frac{cz}{H(z)}\right]^{\frac{1}{3}}
\end{eqnarray}
whose graph as a function of redshift results in the equivalent of BAO Hubble diagram indicating the present day acceleration of the universe (eg., from the WiggleZ survey, \cite{Blake:2011wn, Blake:2011en}).

Upto now it is an open problem of what actually causes the observed acceleration of the universe. There are many possible explanations ranging from a cosmological constant, scalar fields, a  gravity theory different from Einstein general relativity, brane world models, etc.
However, there is no compelling candidate. 
In terms of interpreting the data the simplest approach is to assume a cosmological constant adding just one parameter, $\Omega_{\Lambda}$ which has an equation equation of state $p=-\rho$. Considering a more general equation of state $p=w\rho$ where $w$ is a constant the constraint from observations of the CMB such as, e.g., 
Planck 15 plus astrophysical data gives $w=-1.006\pm 0.045$ \cite{planck15-cosmo}.

If one were to try to connect the observed value of $\Omega_{\Lambda}$
 with the expectation value of a vacuum energy density  a simple estimate shows that there is a huge discrepancy between prediction and observation. 
 Describing vacuum fluctuations in terms of a set of harmonic oscillators at zero point energy yields
 the total energy per unit volume of all oscillators  (e.g. \cite{Perkins:2003pp})
 \begin{eqnarray}
 \epsilon=\frac{\hbar}{4\pi^2}\int \omega_k k^2dk.
 \end{eqnarray}
 Introducing an upper cut-off $k_m$ corresponding to a maximal energy  $E_m$ and considering the relativistic limit leads to $\epsilon=\frac{E_m^4}{16\pi^2(\hbar c)^3}$. A natural energy scale in the problem is the Planck scale $M_P$ for which $\epsilon\sim 10^{121}$ GeV leading to 
 \begin{eqnarray}
 \frac{\rho_{\Lambda}}{\epsilon}\simeq 10^{-121}
 \end{eqnarray}
  for $\rho_{\Lambda}=\Omega_{\Lambda}\rho_c$ and $\Omega_{\Lambda}\sim 0.7$ and $\rho_c\simeq 5$GeV m$^{-3}$.
 This encapsulates the cosmological constant problem (cf., e.g. \cite{Weinberg:1988cp}).

\section{Conclusions}

In these lectures we have provided an overview of the current cosmological model together with observational evidence. Roughly 95 \% of the energy density of the universe is in the "dark" sector for which there is observational evidence but no clear theoretical understanding. There is no  shortage of proposals of models ranging from modifying gravity "by hand", string theory inspired models to various extensions of the standard model of particle physics. However, so far  there is no obvious ("natural") model which stands out.
There will be more data and at higher precision from astrophysical and cosmological observations in the near future.
These can also provide useful additional constraints on particle physics models.

\section{Acknowledgements}

I would like to thank the organizers for inviting me to present these lectures at the CERN Latin-American School of High Energy Physics CLASHEP 2015, which took place in Ibarra, Ecuador.
It was a very interesting school and the organization was marvellous. 
I would like to thank NORDITA in Stockholm, Sweden, for hospitality where part of these lecture notes were written.
Financial support by Spanish Science Ministry grants FPA2015-64041-C2-2-P,  FIS2012-30926 and 
CSD2007-00042 is gratefully acknowledged.

    Funding for the SDSS and SDSS-II has been provided by the Alfred P. Sloan Foundation, the Participating Institutions, the National Science Foundation, the U.S. Department of Energy, the National Aeronautics and Space Administration, the Japanese Monbukagakusho, the Max Planck Society, and the Higher Education Funding Council for England. The SDSS Web Site is http://www.sdss.org/.

    The SDSS is managed by the Astrophysical Research Consortium for the Participating Institutions. The Participating Institutions are the American Museum of Natural History, Astrophysical Institute Potsdam, University of Basel, University of Cambridge, Case Western Reserve University, University of Chicago, Drexel University, Fermilab, the Institute for Advanced Study, the Japan Participation Group, Johns Hopkins University, the Joint Institute for Nuclear Astrophysics, the Kavli Institute for Particle Astrophysics and Cosmology, the Korean Scientist Group, the Chinese Academy of Sciences (LAMOST), Los Alamos National Laboratory, the Max-Planck-Institute for Astronomy (MPIA), the Max-Planck-Institute for Astrophysics (MPA), New Mexico State University, Ohio State University, University of Pittsburgh, University of Portsmouth, Princeton University, the United States Naval Observatory, and the University of Washington.

We acknowledge the use of the Legacy Archive for Microwave Background Data Analysis \newline
(LAMBDA),  part of the High Energy Astrophysics Science Archive Center (HEASARC). \newline
HEASARC/LAMBDA is a service of the Astrophysics Science Division at the NASA Goddard Space Flight Center.

\appendix
\section{A brief introduction to the Friedmann-Robertson-Walker models}
\label{A1}
\setcounter{equation}{0}

In the standard model of cosmology gravity is described by general relativity. As in  special relativity space  and time are united to describe a  space-time. The invariant line element
of special relativity is given by $ds^2=\eta_{\mu\nu}dx^{\mu}dx^{\nu}$ where
\begin{eqnarray}
\eta_{\mu\nu}=\left(
\begin{array}{cccc}
-1 & 0 & 0 &0\\
0&1&0&0\\
0&0&1&0\\
0&0&0&1
\end{array}
\right)
\end{eqnarray}
is the Minkowski metric. In general relativity this is determined  by $ds^2=g_{\mu\nu}dx^{\mu}dx^{\nu}$ for a general metric $g_{\mu\nu}$. Space-time is curved. The dynamics is determined by Einstein's equations,
\begin{eqnarray}
R_{\mu\nu}-\frac{1}{2}g_{\mu\nu}R+\Lambda g_{\mu\nu}=8\pi G T_{\mu\nu}
\end{eqnarray}
where the Ricci tensor is defined by $R_{\mu\nu}=\frac{\partial\Gamma^{\sigma}_{\mu\nu}}{\partial x^{\sigma}}-\frac{\partial\Gamma^{\sigma}_{\mu\sigma}}{\partial x^{\nu}}+\Gamma^{\sigma}_{\rho\sigma}\Gamma^{\rho}_{\mu\nu}-\Gamma^{\sigma}_{\rho\nu}\Gamma^{\rho}_{\mu\sigma}$,
the Christoffel symbols are defined by
 $\Gamma^{\mu}_{\nu\rho}=\frac{g^{\mu\sigma}}{2}\left(\frac{\partial g_{\sigma\rho}}{\partial x^{\nu}}+\frac{\partial g_{\nu\sigma}}{\partial x^{\rho}}-\frac{\partial g_{\nu\rho}}{\partial x^{\sigma}}\right)$ and  $R=g^{\mu\nu}R_{\mu\nu}$ is the Ricci scalar.
Repeated indices in pairs with one contravariant and one covariant are summed over. Greek indices take values between 0 and 3. 
$T_{\mu\nu}$ is the energy-momentum tensor which is conserved, satisfying $\nabla_{\nu}T^{\mu\nu}=0$. 
Speaking in an illustrative way, Einstein's equations encapsulate that geometry determines matter distribution and evolution and vice versa. 
Space-time is curved by the presence of matter. That is why, e.g., the trajectory of light from distant sources is deviated by the sun.
This deflection  angle  is one of the  classical tests of general relativity.

Einstein's equations are very complex. There are no general solutions known. It is always necessary to assume some degree of
symmetry in order to find solutions. The Friedmann-Robertson-Walker solutions are isotropic and homogeneous and are
described by the metric
\begin{eqnarray}
ds^2=-dt^2+a^2(t)\left[\frac{dr^2}{1-kr^2}+r^2\left(d\theta^2+\sin^2\theta d\phi^2\right)\right].
\end{eqnarray}
The parameter $k$ has only three values: $k=0$ for a flat universe, $k=-1$ for an open one and $k=1$ for a closed universe.
In many cases matter can be described by a perfect fluid with 4-velocity $u^{\mu}$ whose 
energy momentum tensor  is given by 
\begin{eqnarray}
T_{\mu\nu}=\left[\rho(t)+P(t)\right]u_{\mu}u_{\nu}+p(t)g_{\mu\nu}
\end{eqnarray}
where $\rho(t)$ and $p(t)$ are the energy density and pressure, respectively, which are only functions of time in 
a Friedmann-Robertson-Walker background.

%%%%%%%%%%%%%%%%%%%%%%%%%%%%%%%%%%%%%%%%%%%%%%%%%%%%%%%%%%%%%%%%%%%%%%%%

\bibliography{references}

\bibliographystyle{hieeetr}

\end{document}